\documentclass[11pt,a4paper]{article}              

\usepackage[margin=20mm]{geometry}
\usepackage{graphicx}
\usepackage{fancyhdr}
\usepackage{amssymb,amsfonts,amsmath,amsthm}
\usepackage{natbib}
\usepackage[margin=1cm]{caption}

\usepackage{hyperref}
\usepackage[english]{babel}
\usepackage[utf8x]{inputenc}
\usepackage[T1]{fontenc}
\usepackage{amsmath}
\usepackage{amsthm}
\usepackage{amstext}
\usepackage{amsfonts}
\usepackage{amssymb}
\usepackage{dsfont}
\usepackage{graphicx}
\usepackage{booktabs}
\usepackage{units}
\usepackage{nth}
\usepackage[capitalize]{cleveref}
\usepackage{paralist}
\usepackage{color}
\usepackage{xfrac}
\usepackage{scalerel}
\usepackage{epstopdf}
\usepackage{hhline}
\usepackage{float}
\usepackage{ragged2e}
\usepackage{times}
\usepackage{parskip}
\usepackage{natbib}
\usepackage{csquotes}
\usepackage{bigints}
\usepackage{relsize}
\usepackage{mathabx}
\usepackage{enumitem}
\usepackage{enumitem}
\usepackage{setspace}
\usepackage{subcaption}
\usepackage{multirow}
\usepackage{array}
\usepackage{arydshln}
\newcolumntype{T}{>{\centering\arraybackslash}m{1.8cm}}
\newcolumntype{L}{>{\centering\arraybackslash}m{3.1cm}}
\newcolumntype{Q}{>{\centering\arraybackslash}m{2.0cm}}

\DeclareMathOperator*{\argminA}{arg\,min}

\usepackage{authblk}
	
\graphicspath{{./},{./Figs/}}

\theoremstyle{plain}

\usepackage{bstnotations}
\usepackage{bm}
\renewcommand{\ve}[1]{\bm{#1}}

\newlength{\dagHeight}	
\newlength{\dagWidth}	
\newlength{\figHeight}	
\newlength{\figWidth}	
\newlength{\beamWidth}	

\begin{document}
\title{Hierarchical Kriging for multi-fidelity aero-servo-elastic
  simulators - Application to extreme loads on wind turbines}

\author[1]{I. Abdallah} \author[2]{C. Lataniotis} 
\author[2]{B. Sudret} 

\affil[1]{Chair of Structural Mechanics,
    ETH Zurich, Stefano-Franscini-Platz 5, 8093 Zurich, Switzerland}

\affil[2]{Chair of Risk, Safety and Uncertainty Quantification,
    ETH Zurich, Stefano-Franscini-Platz 5, 8093 Zurich, Switzerland}

\date{}
\maketitle

\abstract{In the present work, we consider multi-fidelity surrogate modelling to fuse the output of multiple aero-servo-elastic computer simulators of varying complexity. In many instances, predictions from multiple simulators for the same quantity of interest on a wind turbine are available. In this type of situation, there is strong evidence that fusing the output from multiple aero-servo-elastic simulators yields better predictive ability and lower model uncertainty than using any single simulator.
A computer simulator of a physical system requires a high number of runs in order to establish how the model response varies due to the variations in the input variables. Such evaluations might be expensive and time consuming. One solution consists in substituting the computer simulator with a mathematical approximation (surrogate model) built from a limited but well chosen set of simulations output. Hierarchical Kriging is a multi-fidelity surrogate modelling method in which the Kriging surrogate model of the cheap (low-fidelity) simulator is used as a trend of the Kriging surrogate model of the higher fidelity simulator. We propose a parametric approach to Hierarchical Kriging where the best surrogate models are selected based on evaluating all possible combinations of the available Kriging parameters candidates. The parametric Hierarchical Kriging approach is illustrated by fusing the extreme flapwise bending moment at the blade root of a large multi-megawatt  wind turbine as a function of wind velocity, turbulence and wind shear exponent in the presence of model uncertainty and heterogeneously noisy output. The extreme responses are obtained by two widely accepted wind turbine specific aero-servo-elastic computer simulators, FAST and Bladed. With limited high-fidelity simulations, Hierarchical Kriging produces more accurate predictions of validation data compared to conventional Kriging. In addition, contrary to conventional  Kriging, Hierarchical Kriging is shown to be a robust surrogate modelling technique because it is less sensitive to the choice of the Kriging parameters and the choice of the estimation error.
 \\[1em] 

  }

\maketitle

\section{Introduction}

The analysis and design of wind turbines relies on aero-servo-elastic simulators. Aero-servo-elasticity is a term that refers to the coupling of wind inflow, aerodynamics, structural dynamics and controls models. Multiple aero-servo-elastic simulators are available to researchers and engineers, and it is hard to establish that one simulator is better than the other in terms of the predicted output. The attention in the wind energy community has so far been directed towards comparing \citep{Buhl2000, Renewable2006a, Jonkman2008, Jonkman2010, OC42014}, verifying and validating \citep{BuhlJr2001,Simms2001a,Schepers2002} those simulators against each other and against measurements. However, little to no attention has been given to fusing (a.k.a. aggregating, combining) the output from multiple aero-servo-elastic simulators.

In practice we observe that the value of a predicted quantity of interest may vary amongst aero-servo-elastic simulators of the same wind turbine, and it is reasonable to assume that  such variation is of the epistemic type and can be reduced with increasing quantity and quality of the available aero-servo-elastic simulators and simulations output \cite{Kiureghian2009}. The combination of the output from multi-level computer simulators was first proposed by \citet{KennedyOhagan2000} using a  stationary Gaussian process and Bayesian inference where the main assumption is that the output from different simulators levels are auto-regressive, meaning that the output of the higher level simulator is related to the output of the lower level simulator by a regressive coefficient and a discrepency which follows a Gaussian process. \citet{Forrester2007} also used an auto-regressive co-Kriging surrogate modelling approach for multi-fidelity optimization of a generic aircraft wing using one cheap (empirical drag estimation by curve fitting) and one expensive (linearized potential method) flow solver. \citet{Qian2008} adopted a similar approach but employed a fully Bayesian formulation in order to absorb the uncertainty in the model parameters in the prediction. \citet{Kuya2011} used a multi-fidelity co-Kriging surrogate modelling approach where low-fidelity data from computational fluid dynamics (CFD) simulations contribute to improving surrogate models built with limited high-fidelity data from wind tunnel experiments on a single element airfoil equipped with a vortex generator. Recently, \citet{Han2012a} introduced Hierarchical Kriging to predict the mean aerodynamic lift and drag coefficients on a two- dimensional airfoil and a three-dimensional aircraft model using a low-fidelity Euler flow solver and a high-fidelity Navier-Stokes solver. Finally, \citet{LeGratiet2014} proposed a recursive co-Kriging based surrogate modelling approach to predict the output of a complex hydrodynamic simulator of turbulence model for gaseous mixtures from a coarser model by altering the finite-elements mesh.

Most multi-fidelity applications in the literature concern mesh refinement of the same simulator (e.g. in the case of finite element) or describing a single physics phenomenon using more complex approaches (e.g. panel methods versus large eddy simulations in Computational Fluid Dynamics CFD). The reader is referred to \citep{MFCox2016, Godino2016}  for a comprehensive review of multi-fidelity approaches. The aim in this paper is to use multi-fidelity surrogate modelling to combine simulations output from distinct aero-servo-elastic simulators of varying aerodynamics/structural dynamics/turbulent inflow submodels fidelity and complexity. In particular we employ Hierarchical Kriging \citep{Han2012a} to efficiently fuse the extreme flapwise bending moments at the blade root of a large multi-megawatt wind turbine in the presence of heterogeneously noisy output (i.e. the magnitude of noise varies as a function of the input variables) when a low and high-fidelity aero-servo-elastic simulators of the same wind turbine are implemented by two independent engineers (i.e. uncertainty in the modelling and input assumptions are implicitly included). Furthermore, unlike what is found in the literature, we propose a parametric approach to Hierarchical Kriging, where  the best surrogate models are  selected based on evaluating all possible combinations of the available Kriging parameters candidates, namely correlation type, correlation family, correlation isotropy, trend types, hyper-parameters estimation methods and hyper-parameters optimization methods. This, in a way, deals with the model selection problem discussed in \citep{Rasmussen2006}. However here we exploit the architecture of the UQLab software \citep{Marelli2014, UQdoc_09_105} to search for and select the best Krigging surrogate model by parallelizing the process on a computer cluster.

The remainder of this article is organized as follows. In section \ref{sec: casedatafuse} we elaborate the arguments in favour of fusing the output from multiple computer simulators. In section \ref{sec: krigmeta} we revisit the theory of Kriging and in section \ref{sec: hierkrigmeta} we introduce parametric Hierarchical Kriging. In section \ref{sec: appl} we present a real-world engineering application based on aero-servo-elastic simulations of a wind turbine, and compare the performance and accuracy of Hierarchical Kriging and conventional Kriging.

\section{Motivation} \label{sec: casedatafuse}
Increasingly, engineers and researchers use various aero-servo-elastic simulators to analyse and compare the coupled dynamic loads and performance of wind turbines. According to \citet{RanjanGneiting2010} there is a strong empirical evidence that combining output from multiple models results in improved predictive performance. Here we will propose few arguments in favour of fusing the output of multiple aero-servo-elastic simulators. The simulators may exhibit similarities in terms of the input and the underlying  modelling and physics assumptions for wind inflow, aerodynamics, structural dynamics and control systems. In fact, the higher fidelity simulators may simply be an expansion of the lower fidelity simulation model by inclusion of additional physics. In addition, the various aero-servo-elastic simulators may have been coded by the same or cooperating engineers, scientists and research institutes, then calibrated using the same test measurements and the same experts may have given their inputs / reviews / recommendations during the development and validation of the various simulators. Therefore, similar modelling assumptions, biases and possibly gross errors end up being introduced in the various simulators. An important fact is that the various aero-servo-elastic simulators are certified by accredited institutes for use in the industry to design wind turbines. The certification process involves a lengthy verification campaign against measurements. Hence, no particular simulator is necessarily deemed better than the other. Finally, multiple benchmarking studies (\citep{Buhl2000,  BuhlJr2001, Simms2001a, Schepers2002, Renewable2006a, Jonkman2008, Jonkman2010, OC42014}) suggest that the output of the various aero-servo-elastic simulators display similar trends and are generally smooth with respect to small variations in the inputs. Therefore, the implication of the above argumentation strongly points in favour of fusing information from all available aero-servo-elastic simulators rather than discard information from lower fidelity simulators  \citep{Christensen2012}.

\section{Reminder on Kriging} \label{sec: krigmeta}
Let $\ve{x}=\left\lbrace x_1,\ldots,x_d\right\rbrace^T$ be a $d$-dimensional vector of input variables. This input vector is sampled at $n$ distinct locations in the input space $\mathcal{D}_{\ve{x}}$ and the corresponding scalar output are $\left\lbrace \mathcal{Y}^{(i)}=\mathcal{M}\left(\ve{x}^{(i)}\right), \mathbin{} i=1,\ldots,n\right\rbrace$, where $\mathcal{M}$ is the computer simulator of interest. The $n$ distinct input samples are collected in $\bm{\mathcal{X}} = \left(\ve{x}^{(1)},\ldots,\ve{x}^{(n)}\right)$ and the output $\boldsymbol{\mathcal{Y}} = \left\lbrace \mathcal{Y}^{(1)},\ldots,\mathcal{Y}^{(n)}\right\rbrace^T$. For a given simulator $\left\lbrace \mathcal{M}_l, \mathbin{} l=1,\ldots,s \right\rbrace$ the vector of output is $\boldsymbol{\mathcal{Y}}_l$, $s$ being the total number of available simulators of varying complexity and fidelity. We define $\mathcal{M}^{K}$ the Kriging surrogate model of the simulator $\mathcal{M}$.

A Kriging surrogate model approximates the output from a computer experiment which is a collection of pairs of input and responses from runs of a computer simulator\cite{Sacks1989,Santner2003}. Suppose that for a given input $\mathcal{X}$ the scalar output $\mathcal{Y}=\mathcal{M}\left(\ve{x}\right)$ is a realization of a random variable $Y$ that is normally distributed with mean $\mu$ and variance $\sigma^2$. We collect all such random variables at $n$ input samples in a random vector $\ve Y$:
\begin{equation}
	\ve Y = \left\lbrace Y^{(1)} \quad Y^{(2)} \quad Y^{(3)} \quad \ldots \quad Y^{(n)} \right\rbrace^T
\end{equation}
\citet{Rasmussen2006} define a Gaussian process as an infinite collection of random variables, any finite number of which having a multivariate Gaussian distribution. Each random variable $Y^{(i)}$ is normally distributed and hence $\ve Y$ is a Gaussian random vector with a $n\times 1$ vector of means $\ve 1\mu$ and an $n\times n$ covariance matrix $\ve C$: $\ve Y\sim \mathcal{N}_n\left(\ve 1\mu, \ve C\right)$. Kriging is a stochastic interpolation technique which assumes that the model output is a realization of a Gaussian process \citep{Cressie1993, Santner2003, Rasmussen2006}:
 \begin{equation}
  \begin{split}
	\mathcal{M}\left(\ve{x}\right) \approx \mathcal{M}^{K}\left(\ve{x}\right) &= \mu\left(\ve{x}\right) + \sigma^2 Z(\ve{x})  \\
	&= \sum_{j=1}^{P}\beta_jf_j(\ve{x}) + \sigma^2 Z\left(\ve{x}\right)\\
  \end{split}
 	\label{eqn: gaussprocess}
\end{equation} 
where $\mu\left(\ve{x}\right)$ is the mean of the Gaussian process, also known as the trend which consists of the regression coefficients $\left\lbrace\beta_j, j=1,\ldots,P\right\rbrace$ and the basis functions $\left\lbrace f_j, j=1,\ldots,P\right\rbrace$, while the systematic lack-of-fit (departure from the expected value) is modelled by $\sigma^2Z\left(\ve{x}\right)$ which effectively "pulls" the Kriging function through the observed simulations output by quantifying the correlation of nearby points. $\sigma^2$ is the constant variance of the Gaussian process, and $Z\left(\ve{x}\right)$ is a zero-mean, unit-variance stationary Gaussian process. $Z(\ve{x})$ is fully determined by the correlation function $R\left(\ve{x},\ve{x}'\right)$ between two distinct points $\left(\ve{x},\ve{x}'\right)$ in the input space:
\begin{equation}
	R\left(\ve{x},\ve{x}'\right)=R\left(\ve{x}-\ve{x}' \mathrel{\stretchto{\mid}{3ex}}  \ve \theta\right)
 	\label{eqn: univarCovar}
\end{equation}
where the hyper-parameters $\ve \theta$ are to be computed. From the Design of Experiments (DOE) $\boldsymbol{\ve{x}}$, one can build the correlation matrix with terms  $\ve R_{ij} = R\left(\ve{x}^{(i)}-\ve{x}^{(j)} \mathrel{\stretchto{\mid}{3ex}}  \ve \theta\right)$ representing the correlation between the outputs at the input combinations $\ve{x}^{(i)}$ and $\ve{x}^{(j)}$. Table \ref{tab:corrfamilieslist} lists the most common correlation functions that can be found in the literature \citep{Santner2003,Rasmussen2006}. Different correlation families result in different levels of smoothness for the associated Gaussian process. 



\begin{table}[!ht]
\small
  \caption{Most common correlation families used in Kriging. $h=\lvert\lvert\ve{x}-\ve{x}'\rvert\rvert$.}
  \label{tab:corrfamilieslist}
  \centering
\begin{tabular}{ p{3cm} p{7cm} }
      \hhline{==}
      Correlation family & Formula\\
          \midrule
    Gaussian       &	$R\left(h \mathrel{\stretchto{\mid}{3ex}} \theta_q \right)=\exp\left(-\sum_{q=1}^d\left(\frac{h}{\theta_q}\right)^2\right)$			\\
    Exponential    & $R\left(h \mathrel{\stretchto{\mid}{3ex}} \theta_q \right)=\exp\left(-\frac{\lvert h \rvert}{\theta_q}\right)$			\\
    Mat\'ern-3/2   &	 $R\left(h \mathrel{\stretchto{\mid}{3ex}} \theta_q \right)=\left(1 + \frac{\sqrt{3}\lvert h \rvert}{\theta_q}\right)\exp\left(-\frac{\sqrt{3}\lvert h \rvert}{\theta_q}\right)$				\\
	Mat\'ern-5/2   &		$R\left(h \mathrel{\stretchto{\mid}{3ex}} \theta_q \right)=\left(1 + \frac{\sqrt{5}\lvert h \rvert}{\theta_q}+\frac{5h^2}{3\theta_q^2}\right)\exp\left(-\frac{\sqrt{5}\lvert h \rvert}{\theta_q}\right)$			\\
    Linear			    &  $R\left(h \mathrel{\stretchto{\mid}{3ex}} \theta_q \right)=\max\left(0,1 - \frac{\lvert h \rvert}{\theta_q}\right)$ \\

      \hhline{==}
\end{tabular}
\end{table}

When $d>1$, the correlation functions could be anisotropic and separable, which reads \citep{Sacks1989}
 \begin{equation}
	R\left(\ve{x},\ve{x}' \mathrel{\stretchto{\mid}{3ex}} \ve \theta\right) =\prod_{q=1}^d R\left(x_q,x'_q \mathrel{\stretchto{\mid}{3ex}} \theta_q\right)
\end{equation}
or anisotropic ellipsoidal \citep{Rasmussen2006}
\begin{equation}
	R\left(\ve{x},\ve{x}' \mathrel{\stretchto{\mid}{3ex}} \ve \theta\right) = R\left(h\right) , h=\sqrt{\sum_{q=1}^d\left(\frac{x_{q}-x'_{q}}{\theta_q}\right)^2}
\end{equation}
In these equations $x_{q}$ and $x'_{q}$ are the $q$-th coordinate of the $\ve{x}$ and $\ve{x}'$.

Generally speaking, a correlation function is called isotropic when it has the same behaviour over all dimensions, meaning that the hyper-parameters in the previous two equations become $\theta_q=\theta$ for every dimension in the input space $\mathcal{D}_{\ve{x}}$  \citep{Marelli2014, UQdoc_09_105}. 

To fully define the distribution of $\ve Y$ we need to estimate the values of $\beta_j$, $\sigma^2$ and $\theta_q$ in every dimension $q=1,\ldots,d$. This is achieved by solving an optimization problem that differs depending on the estimation method that is used. The estimation methods that are available include the maximum likelihood estimation (MLE) and the leave-one-out cross-validation method (CV). In the Maximum Likelihood method the parameters $\beta_j$, $\sigma^2$ and $\theta_q$ are chosen such that the likelihood of the observed output data $\boldsymbol{\mathcal{Y}}$ is maximized. Given that $\ve Y$ is a Gaussian vector, maximizing the log-likelihood of the multivariate normal distribution one gets the following estimates of the Gaussian process variance $\hat{\sigma}^{2}$ and Kriging trend coefficients $\hat{\boldsymbol{\beta}}$ that are known as the generalized least-squares estimates (for proof and more details the reader is referred to  \citep{Santner2003,Marrel2008PhD,DubourgThesis}:
\begin{align}
	& \hat{\sigma}^{2}(\hat{\boldsymbol{\theta}}) = \frac{(\ve {\mathcal{Y}}-\ve {F}\hat{\ve {\beta}})^T\ve {R}^{-1}(\ve {\mathcal{Y}}-\ve {F}\hat{\ve {\beta}})}{n}\\
	& \hat{\boldsymbol{\beta}}(\hat{\boldsymbol{\theta}})= \left(\ve {F}^T\ve {R}^{-1}  \ve {F}\right)^{-1}\ve {F}^T \ve {R}^{-1} \ve {\mathcal{Y}}
 	\label{eqn: beta}
\end{align}
where $\ve R$ is the correlation matrix and $\ve F$ is the information matrix (see Eq.~(\ref{eqn: krigmatrices})  below). However, $\hat{\sigma}^{2}$ and $\hat{\boldsymbol{\beta}}$ are dependent on the value of the hyper-parameters $\hat{\boldsymbol{\theta}}$ which are calculated by solving the optimization problem \citep{Santner2003}: 
\begin{equation}
	\hat{\boldsymbol{\theta}}= \argminA_{\mathcal{D}_{\boldsymbol{\theta}}} \left[\frac{1}{2}\log\left(\det\left(\ve {R}\right)\right) + \frac{n}{2}\log\left(2\pi\sigma^2\right) + \frac{n}{2} \right]
 	\label{eqn: thetaoptML}
\end{equation}

If the leave-one-out cross-validation \citep{Santner2003,Bachoc2013b} is used, the optimal parameters are determined through a minimization which reads:
\begin{equation}
	\hat{\boldsymbol{\theta}}= \argminA_{\mathcal{D}_{\boldsymbol{\theta}}} \left[\sum_{i=1}^n\left(\mathcal{M}\left(\ve{x}^{(i)}\right)-\mu_{Y,(-i)}\left(\ve{x}^{(i)}\right)\right)^2\right]
 	\label{eqn: thetaoptCV}
\end{equation}
where $\mu_{Y,(-i)}\left(\ve{x}^{(i)}\right)$ denotes the mean Kriging predictor that was trained on all the data except $\left(\ve{x}^{(i)},\mathcal{Y}^{(i)}\right)$. For solving the optimization problems described in Eq.~ (\ref{eqn: thetaoptML}) or (\ref{eqn: thetaoptCV}) there are trade-offs for choosing some local, usually gradient-based methods such as the quasi-Newton Broyden-Fletcher-Goldfarb-Shanno (BFGS) algorithm \citep{Goldfarb1970,Fletcher1970,Shanno1970,Nocedal1980} and its modifications \cite{Byrd1999}, or global methods, usually evolutionary algorithms such as genetic algorithms \cite{Goldberg1989} and differential evolution algorithms \citep{Storn1997,Deng2013}. The best optimization algorithm is problem dependent and in many cases not known a priori \cite{SchoebiIJUQ2015}.

Next we derive the analytical expressions for the Kriging predictor at a new point $\ve{x}^{(*)}$ in the input space. Let $y^{(*)}$ be the unobserved output at $\ve{x}^{(*)}$ which we would like to predict, then the vector of observations augmented by $y^{(*)}$ becomes:
\begin{equation}
\tilde{\boldsymbol{\mathcal{Y}}}=\left\lbrace\boldsymbol{\mathcal{Y}}^T \quad y^{(*)}\right\rbrace^T  
\end{equation}
We denote $\ve r$ the vector of cross-correlations between the point $\ve{x}^{(*)}$ where the prediction is to be performed and each of the points of the experimental design $\ve{x}^{(i)}$ where observations $\mathcal{M}\left(\ve{x}^{(i)}\right)$ already exist:
\begin{equation}
	\ve r=\left\lbrace R\left(\ve{x}^{(*)}-\ve{x}^{(1)} \mathrel{\stretchto{\mid}{3ex}}  \ve \theta\right),\ldots,R\left(\ve{x}^{(*)}-\ve{x}^{(n)} \mathrel{\stretchto{\mid}{3ex}}  \ve \theta\right) \right\rbrace^T
 	\label{eqn: rcorrvect}
\end{equation}
The augmented correlation matrix then becomes:
\begin{equation}
\tilde{\ve R}=
\begin{bmatrix}
	\ve R & \ve r \\
	\ve r^T & 1
\end{bmatrix}
 	\label{eqn: corramtaugm}
\end{equation}
and the Kriging predictor (predicted response conditional on the observed output of the simulations) at a new point $\ve{x}^{(*)}$ is a conditional Gaussian variable $Y^{(*)} \mathrel{\stretchto{\mid}{3ex}}  \boldsymbol{\mathcal{Y}}$ with mean $\mu_{Y^{(*)}}$ and Kriging prediction variance $\sigma^2_{Y^{(*)}}$:
\begin{equation}
\begin{split}
	\mu_{Y^{(*)}} &= \mathbb{E}\left[Y^{(*)} \mathrel{\stretchto{\mid}{3ex}} \boldsymbol{\mathcal{Y}}\right]\\
	&= \hat{\boldsymbol{\beta}}\boldsymbol{f}^T\left(\ve{x}^{(*)}\right)  +  \boldsymbol{r}^T \boldsymbol{R}^{-1} \left(\boldsymbol{\mathcal{Y}} - \boldsymbol{F}\hat{\boldsymbol{\beta}}\right)\\
\end{split}
 	\label{eqn: meanpred}
\end{equation}
\begin{equation}
	\begin{split}
	\sigma^2_{Y^{(*)}} &=  \mathrm{Var}\left[\hat{Y}_* \mathrel{\stretchto{\mid}{3ex}}  \boldsymbol{\mathcal{Y}}\right]\\ 	
	 &=   \hat{\sigma}^{2} \left[1-\boldsymbol{r}^T \boldsymbol{R}^{-1}\boldsymbol{r} + \boldsymbol{u}^T (\boldsymbol{F}^T\boldsymbol{R}^{-1} \boldsymbol{F})^{-1} \boldsymbol{u}\right]
\end{split}
 	\label{eqn: varpred}
\end{equation}
where $\ve {u}$ and the information matrix $\ve {F}$ are given by:
\begin{gather}
	\ve {u} = \ve {F}^T\ve {R}^{-1}\ve {r} - \ve {f}\\
	\ve {F} = \left[ f_j(\ve{x}^{(i)})\right]=
	\begin{bmatrix}
	f_1(\ve{x}^{(1)}) & \dots & f_P(\ve{x}^{(1)}) \\
    \vdots													  \\
	f_1(\ve{x}^{(n)}) & \dots  & f_P(\ve{x}^{(n)}) \\
	\end{bmatrix}
 	\label{eqn: krigmatrices}
\end{gather}
In the case of simple Kriging, $\mu$ is assumed to be a known constant. In the case of ordinary Kriging, $\mu$ is assumed to be an unknown constant.  In the case of universal Kriging $\mu(\ve{x})$ is cast as $\ve f^T\ve \beta=\sum_{j=1}^{P}\beta_jf_j(\ve{x})$, which is a linear combination of $P$ unknown regression coefficients $\beta_j$ and a set of preselected basis functions $f_j(\ve{x})$ usually predefined as polynomial functions of degree $\kappa$. Essentially, ordinary Kriging is universal Kriging with one basis function $f_1=1$ and one unknown parameter $\beta_1=\mu$. 

The Kriging prediction variance depends on the quantity of available knowledge (observed output). In other words, the uncertainty in the output prediction is purely epistemic and due to a lack of knowledge at specific input $\ve{x}^{(*)}$. The variance of $y^{(*)}$ is zero whenever $\ve{x}^{(*)}=\ve{x}^{(i)}$ because we know exactly the observed output at each of the training points $\ve{x}^{(i)}$ and there is no error term in the stochastic process model \cite{Santner2003}. When the distance between two input points $\ve{x}^{(i)}$ and $\ve{x}^{(j)}$ is small, the correlation among the responses is assumed to be relatively high. In terms of prediction at unobserved input point $\ve{x}^{(*)}$, this implies that nearby sampled points will have more impact on the interpolation of predicted values than points far away. 

When the outputs of the computer experiments contain noise, the Kriging model should regress the data (instead of interpolating) in order to generate a smooth fit. This is known as Gaussian process regression \cite{Goldberg1998}. The Kriging thus amounts to conditioning  $Y^{(*)}$ on the noisy observations $ \boldsymbol{\mathcal{Y}} + \epsilon$. Kriging predictor mean $\mu_{\hat{Y}_*}(\ve{x}^{(*)})$ and variance $\sigma^2_{\hat{Y}_*} (\ve{x}^{(*)})$  are given by Eqs.~(\ref{eqn: meanpred}) and (\ref{eqn: varpred}), respectively by replacing the correlation matrix $\ve R$ with $\ve R+\sigma_{\epsilon}^2 \ve I_n$, where the constant $\sigma_{\epsilon}^2$ is the estimated variance of the noise term $\epsilon$ and $\ve I_n$ is an $n\times n$ identity matrix. A different $\sigma_{\epsilon}^2$ value can be added to each diagonal element of $\ve R$ when the noise is not constant as a function of the input (heterogeneous noise variances) \citep{Picheny2013a}.

\section{Hierarchical Kriging} \label{sec: hierkrigmeta} 
\subsection{Problem definition}
We consider $s$ aero-servo-elastic simulators of varying fidelity and computational complexity. The approach adopted here is to build Hierarchical Kriging surrogate models by building a Kriging surrogate of the cheap (low-fidelity) simulator first and then using it as a trend of the Kriging surrogate of the higher fidelity simulator. For any given level $1 \leq l \leq s$, the Hierarchical Kriging predictor at an unobserved point $\ve{x}^{(*)}$ can be written as \cite{Han2012a}:
\begin{equation}
\mu_{Y_l^{(*)}}= \hat{\ve {\beta}} \mu_{Y_{l-1}^{(*)}}  +  \ve {r}^T \ve {R}^{-1} (\ve {\mathcal{Y}}_l - \ve {F}\hat{\ve {\beta}})
 	\label{eqn:HK}
\end{equation} 
where $\ve {\mathcal{Y}}_l$ is the vector of output from computer simulator $l$, $\hat{\ve {\beta}}$ is a vector of regression (scaling) factors having a similar expression as in Eq.~(\ref{eqn: beta}). $\mu_{Y_{l-1}^{(*)}} $ is the Kriging predictor of the output of computer simulator $l-1$ and the expression $\ve {R}^{-1} (\ve {\mathcal{Y}}_l - \ve {F}\hat{\ve {\beta}})$ depends only on the observed output at level $l$. We note the similarity of the expressions in Eqs.~(\ref{eqn: meanpred}) and (\ref{eqn:HK}) where $\mu_{Y_{l-1}^{(*)}} $ replaces $\boldsymbol{f}$ in Hierarchical Kriging. The variance of the Hierarchical Kriging predictor is analogous to the expression of Universal Kriging and is given by  \cite{Han2012a}:
\begin{equation}
	\sigma^2_{\hat{Y}_{l}^{(*)}} =   \hat{\sigma}^{2} \left[1-\boldsymbol{r}^T \boldsymbol{R}^{-1}\boldsymbol{r} + \left[\ve {F}^T\ve {R}^{-1}\ve {r} - \mu_{Y_{l-1}^{(*)}} \right]^T (\boldsymbol{F}^T\boldsymbol{R}^{-1} \boldsymbol{F})^{-1} \left[\ve {F}^T\ve {R}^{-1}\ve {r} - \mu_{Y_{l-1}^{(*)}}  \right]\right]
 	\label{eqn: varpredHK}
\end{equation} 

The formulation and implementation of Hierarchical Kriging presented here offers several advantages. It is computationally cheap: we perform one Kriging estimation per fidelity level of a computer simulator. It is flexible in terms of how the Kriging models are derived: since we perform one Kriging estimation per fidelity level of a computer simulator, each of the Kriging models may be defined by a different correlation kernel. In addition, we use a one-step optimization process to determine the Kriging hyper-parameters for every Kriging model. This approach does not make any assumption on the correlation amongst computer simulators, and does not require that input $\boldsymbol{\ve{x}}_l$ be a subset of $\boldsymbol{\ve{x}}_{l-1}$. Furthermore, unlike conventional co-Kriging \citep{Sacks1989,KennedyOhagan2000}, Hierarchical Kriging does not require the modelling of a large cross-covariance matrix, which reduces the error in estimating the hyper-parameters of the Kriging models. In addition, with Hierarchical Kriging we have $s$ covariance matrices to invert, each of which is smaller, less expensive and potentially less ill-conditioned than one large covariance matrix and the estimation of the correlation Kernel parameters can be performed separately  \citep{LeGratiet2014}. Finally, Hierarchical Kriging entails very little (or no) modifications to an existing Kriging code if the latter is sufficiently modular.

\subsection{Analytical example}
\begin{figure}[!ht]
  \begin{subfigure}[b]{0.5\linewidth}
    \centering
  \includegraphics[scale=0.5]{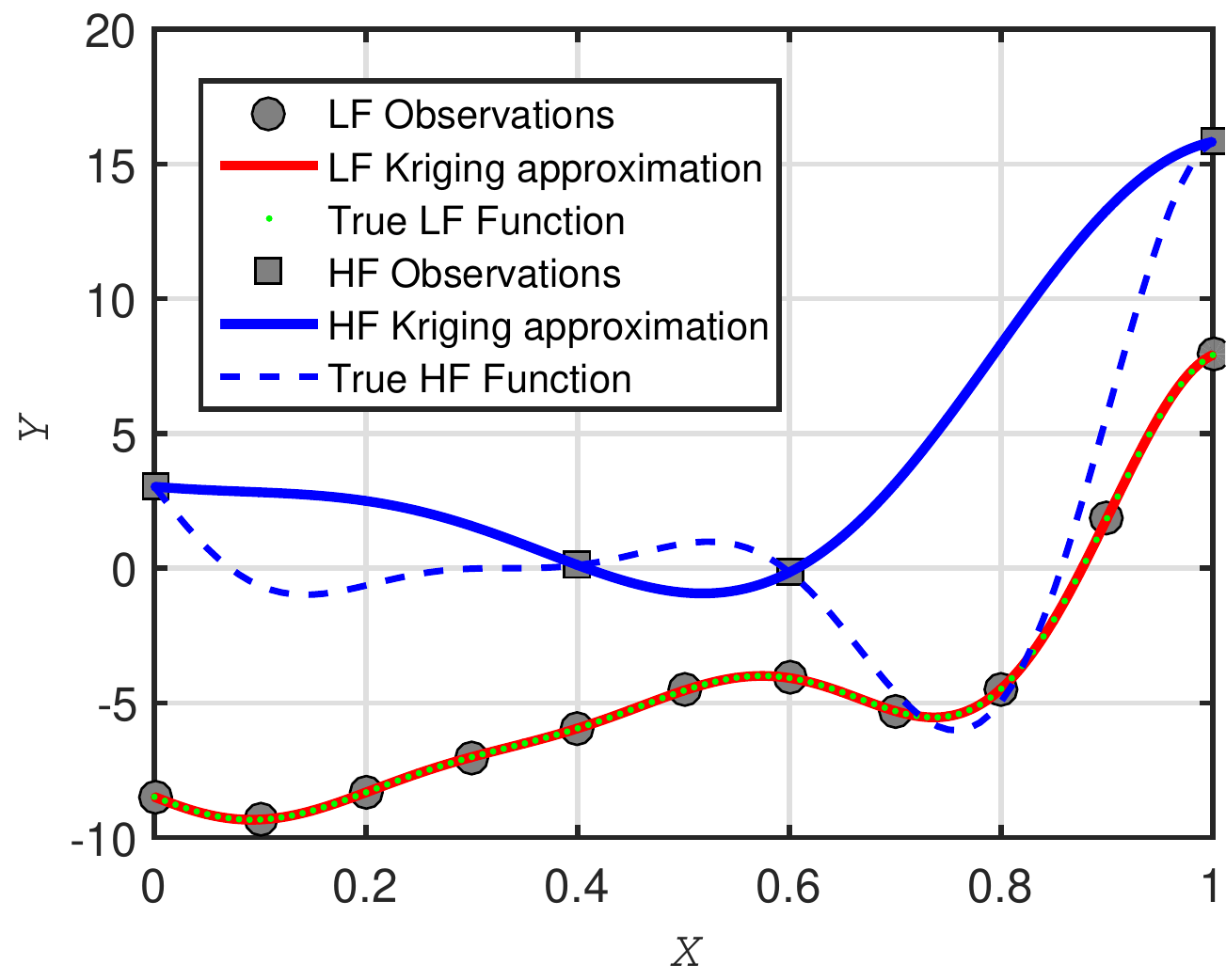}
\caption{}
  \label{fig:coKrig1}
    \vspace{0ex}
  \end{subfigure}
  \begin{subfigure}[b]{0.5\linewidth}
    \centering
  \includegraphics[scale=0.5]{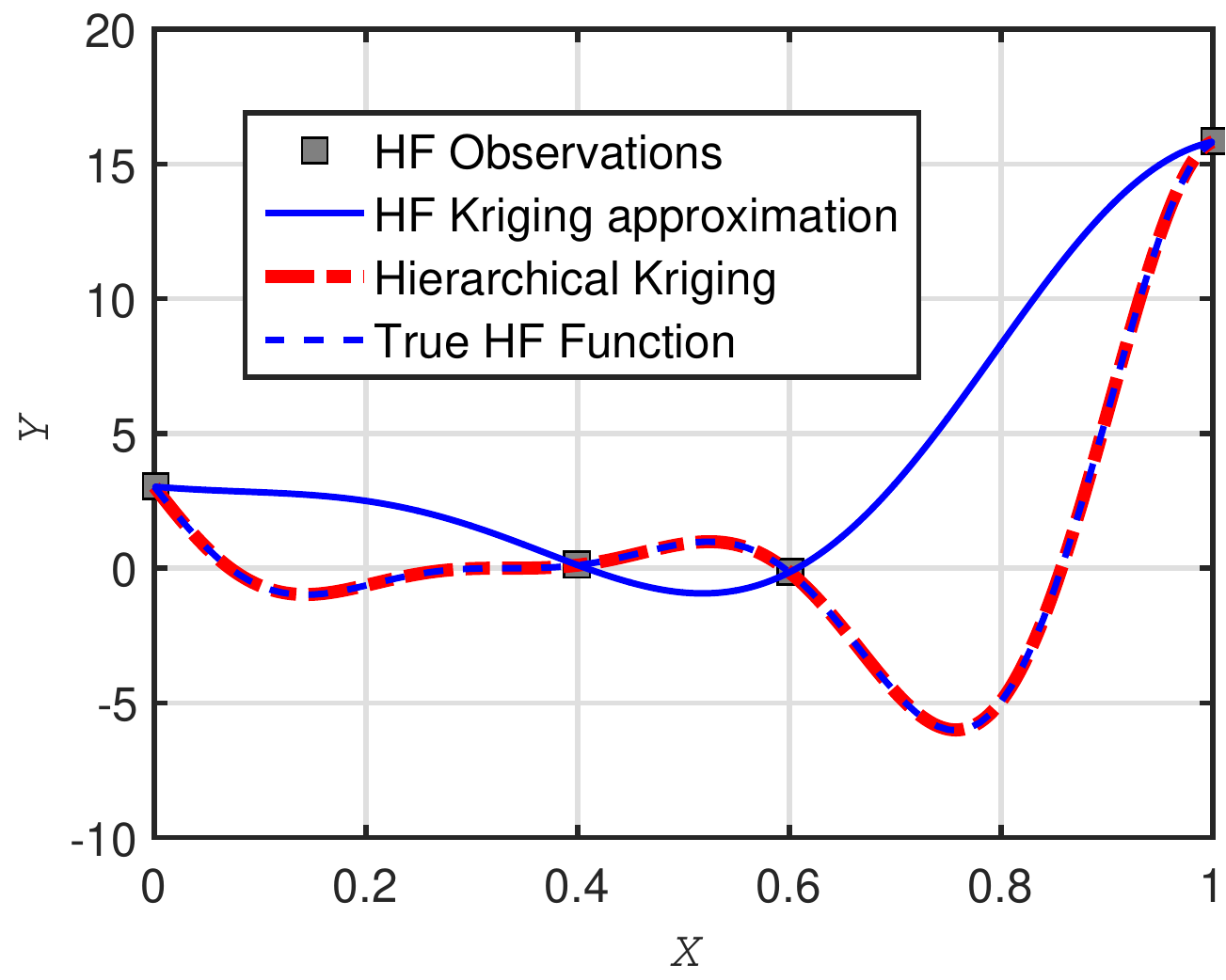}
\caption{}
  \label{fig:DOE2}
    \vspace{0ex}
  \end{subfigure} 
  \caption{(a) Ordinary Kriging approximation of the \emph{low} (LF) and \emph{high} fidelity (HF) analytical functions with a Gaussian correlation kernel. (b) Comparison of conventional Kriging and Hierarchical Kriging approximations of the \emph{high} fidelity analytical function. The Hierarchical Kriging $\ve \hat{\beta}=\left\lbrace 1.99 \right\rbrace$, $\ve \hat{\theta}=9.85$ and $\hat{\sigma}^2=77.89$.}
  \label{fig:coKrig2}
\end{figure}

Hierarchical Kriging is demonstrated on a one dimensional analytical function using UQLab \citep{Marelli2014, UQdoc_09_105}. Suppose that the high- and low-fidelity analytical functions are of the form \cite{Forrester2007}:
\begin{align}
	& y_{hf}(x) = (6x-2)^2 \sin(12x-4) \\
	& y_{lf}(x) = Ay_{hf}(x) + B(x-0.5) - C
\end{align}  
where the input parameter $x$ varies over $\left[0,1\right]$. The constants are set to $A=0.5$, $B=10$ and $C=-5$. The experimental design of the low-fidelity model is $D_1 = \{0,0.1,0.2,\ldots,1\}$, and the experimental design of the high-fidelity model is $D_2 = \{0,0.4,0.6,1\}$. The results are compared in Figure ~\ref{fig:coKrig2} which shows how the Kriging approximation using four observations of the high-fidelity function has been significantly improved using extensive sampling from the low-fidelity function.

\subsection{Parametric Hierarchical Kriging}  \label{sec: compoptkrigmodels}
We often find in the literature that the Kriging surrogate models of computer simulators are calculated based on a single choice of  the correlation kernel, trend type and hyper-parameters estimation method (Kriging parameters) without much explanation given. However, for most engineering problems, the set of Kriging parameters that yield the best surrogate model is not known a priori and may be affected by the choice of the design of experiments. A search through an ensemble of Kriging parameters may prove advantageous to protect against a poor or lucky choice of a surrogate. This section deals with the application of high performance computing to UQLab \citep{Marelli2014, UQdoc_09_105}, with the overall goal of searching for the best Hierarchical Kriging model which, is selected based on evaluating all possible combinations of the available Kriging parameters candidates, namely: (1) correlation type, (2) correlation family, (3) correlation isotropy, (4) trend types, (5) hyper-parameters estimation methods and (6) hyper-parameters optimization methods. Table~\ref{tab:paramstudy} shows the Kriging parameters, resulting in $600$ unique combinations upon which the best Hierarchical Kriging model is selected. The hyper-parameters estimation methods are the Maximum Likelihood (MLE) and the Cross-Validation (CV) estimation methods with a Leave-One-Out option. Gradient-based optimization method BFGS \citep{Byrd1999,Nocedal1980} is considered as one of the options for solving the hyper-parameters optimization problem. Alternatively, the global optimization method Hybrid Genetic Algorithm (HGA \cite{Goldberg1989}), which is a hybrid approach where the final solution of the genetic algorithm is used as a starting point of the gradient method that was previously mentioned. Furthermore, the Hybrid Self-Adaptive Differential Evolution (HSADE \cite{Deng2013}) method is considered, where the optimal solution that is obtained is then refined using BFGS. For each combination of the Kriging parameters a Kriging surrogate model with variable nugget effect is first fitted to all the noisy low-fidelity simulations. The low-fidelity Kriging model is then used as a model trend to fit a Hierarchical Kriging model with variable nugget effect to the noisy high-fidelity simulations. A subset of the high-fidelity simulations are used as a training sample to build the Hierarchical Kriging surrogate model while the remaining output observations are used as validation points. Then each of the ensuing 600 surrogate models are used to predict the high-fidelity simulator output at the validation points and compare the predictions with the true values. The combination of the Kriging parameters resulting in the Hierarchical Kriging model with the lowest estimation error is deemed the best possible surrogate model candidate of the high-fidelity simulations given the available experimental design. Given the architecture of UQLab, the selection of the best surrogate model was done by parallelizing the search on a cluster computer.

\begin{table}[!ht]
\small
  \caption{Various combinations of Kriging options to set-up Hierarchical Kriging and conventional Kriging surrogate models of the high-fidelity simulations.}
  \label{tab:paramstudy}
  \centering
\begin{tabular}{T T T  T  p{3cm}  p{3.3cm} }
      \hhline{======}
      Correlation type & Correlation family &  Correlation isotropy & Trend type & Hyper-parameters estimation method  & Hyper-parameters optimization method\\
          \midrule
    Separable 		& Gaussian       & True 		    &		Ordinary							 & MLE		    & HSADE				\\
    Ellipsoidal    & Exponential  &  False         &     Polynomial (order 1-4)    & CV   			& Hybrid GA			\\
                         	& Mat\'ern-3/2   &   					&													 &					& BFGS					\\
							& Mat\'ern-5/2   &                    &           										 &					&	    						\\
                         	& Linear			    &                    &          						 				 &					&								\\
      \hhline{======}
\end{tabular}
\end{table}

\subsection{Error estimation}  
The predictive coefficient ($Q_2$) and the Maximum Absolute Error ($MAE$) were used to measure the accuracy of the Hierarchical Kriging surrogate models. $Q_2$ is a global metric of the accuracy of the surrogate model \citep{Marrel2012}. It is computed from a validation set and is given by:
\begin{equation}
Q_2=1-\frac{\sum_{i=1}^{n_v}\left[\mathcal{Y}^{(i)} - \mathcal{M}^{K}\left(\ve{x}^{(i)}\right)\right]^2}{\sum_{i=1}^{n_v}\left[\mathcal{Y}^{(i)} - \mu_{\mathcal{Y}}\right]^2}, \quad \mu_{\mathcal{Y}}=\frac{1}{n}\sum_{i=1}^{n_v}\mathcal{Y}^{(i)}
\end{equation}
where $n_v$ is the size of the validation set, and $\mu_{\mathcal{Y}}$ is the mean of the computer simulator response for the validation set. Values of $Q_2$ close to 1 indicate a good fit. $MAE$ on the other hand is a local error metric: 
\begin{equation}
MAE = \frac{\max\left(\mathrel{\stretchto{\mid}{3ex}} \mathcal{Y}^{(1)} - \mathcal{M}^{K}\left(\ve{x}^{(1)}\right)\mathrel{\stretchto{\mid}{3ex}} ,\ldots,\mathrel{\stretchto{\mid}{3ex}} \mathcal{Y}^{(n_v)} - \mathcal{M}^{K}\left(\ve{x}^{(n_v)}\right)\mathrel{\stretchto{\mid}{3ex}} \right)}{\mathrel{\stretchto{\mid}{3ex}}\max\left(\mathcal{Y}\right)-\min\left(\mathcal{Y}\right)\mathrel{\stretchto{\mid}{3ex}}}
\end{equation}
$MAE$ is normalized by the output range in the validation set. A large value of $MAE$ could indicate that the Hierarchical Kriging surrogate model is inaccurate in a particular region of the design space.

\section{Application to Extreme Loads on a Wind Turbine} 
\begin{figure}[h!]
  \centering
  \includegraphics[scale=0.35]{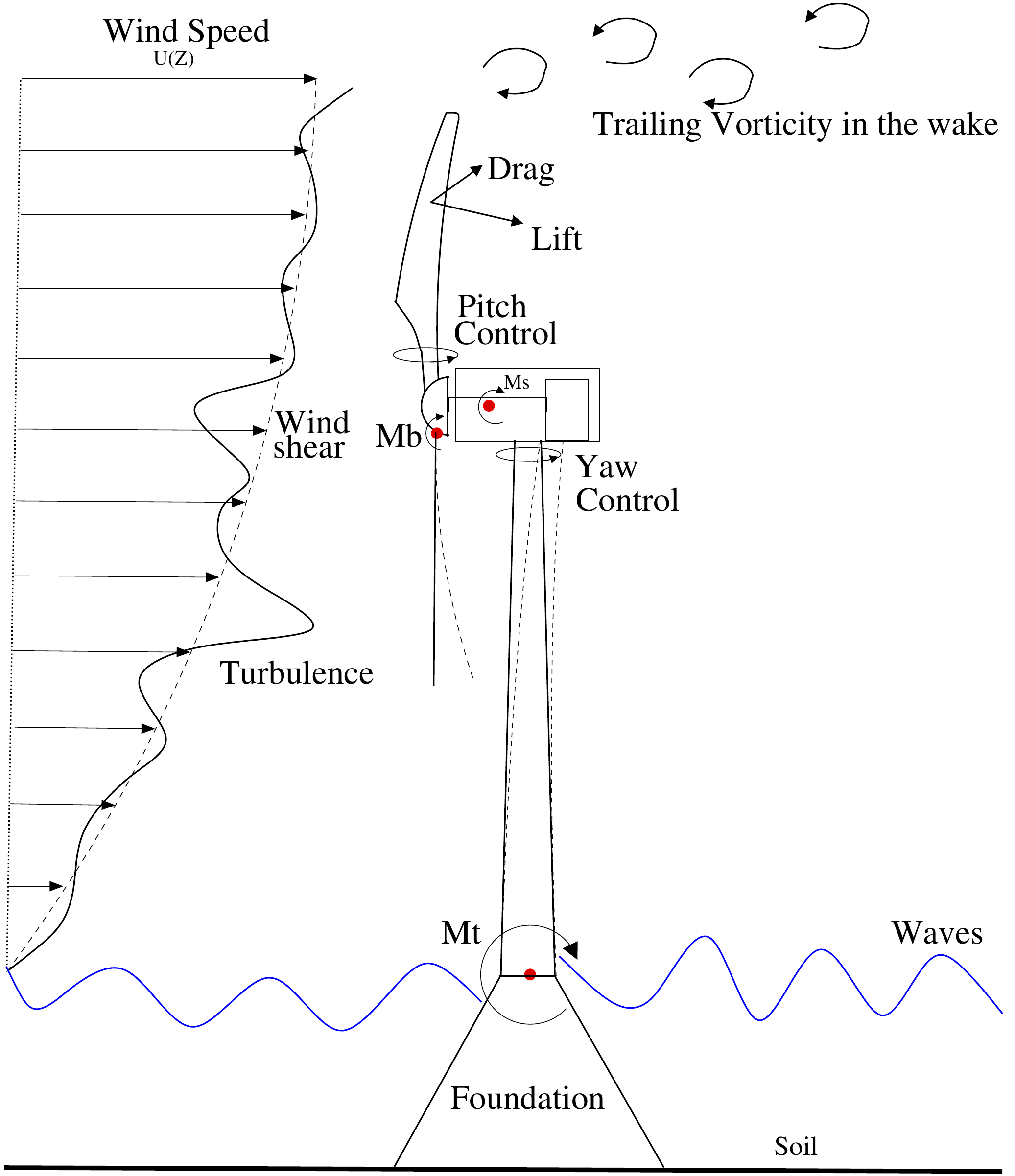}
  \caption{Scheme of a wind turbine: $M_b$ is the flapwise bending moment at the blade root. $U(Z)$ is the mean wind velocity at height $Z$, vertical wind shear is depicted by the dotted grey line and turbulence by thick black line.}
  \label{fig:wtg}
\end{figure}

\label{sec:
  appl} A common practice during the design of a wind turbine is to
generate a significant number of stochastic simulations for various
operating and environmental conditions. In this section we show how the
wind turbine loads simulations are used to demonstrate a real-world
engineering application of Hierarchical Kriging using two
aero-servo-elastic computer simulators, FAST \cite{Jonkman2005}
(low-fidelity) and Bladed \cite{Bladedtheory2003} (high-fidelity). The
FAST and Bladed simulators were implemented and run by two independent
engineers (one of whom is the first author of this paper), and as a
result uncertainty in the modelling and input assumptions are implicitly
included. Hierarchical Kriging is used to fuse the extreme flapwise
bending moment at the blade root ($M_b$), see Figure~\ref{fig:wtg}. The
Hierarchical Kriging surrogate models are then compared to conventional
Kriging of the high-fidelity simulators.

\subsection{Description of the Wind Turbine}\label{sec: thewindturbine}
The simulations in Bladed and FAST considered an onshore utility-scale variable-pitch and variable-speed upwind wind turbine that has a $110~m$ rotor diameter and $2~MW$ rated power. The wind turbine is erected on a $90~m$ tower. We list some of the more important properties of the simulated wind turbine in Table~\ref{tab:WTGProps}.

\begin{table}[h!]
  \caption{Wind turbine properties.}
  \label{tab:WTGProps}
  \centering
    \begin{tabular}{p{5cm}  p{8cm}}
      \hhline{==}
      Number of blades & $3$\\
      Rotor diameter & $11~0m$\\
      Hub height & $90~m$\\
      Rated power & $2~MW$\\
	  Cut-in wind velocity & $4~m/s$\\
      Cut-out wind velocity & $25~m/s$\\
      Control & Variable Speed, Collective Pitch, Active Yaw\\
      Variable speed & from cut-in to cut-out wind velocity\\
      Variable pitch & from cut-in to cut-out wind velocity\\
      Rated wind velocity & $10.5~m/s$\\
      Cut-in and rated RPM & $8.5-13~RPM$\\ 
      \hhline{==}
    \end{tabular}
\end{table}

\subsection{Design of experiments}
\begin{figure}[h!]
  \begin{subfigure}[b]{0.5\linewidth}
    \centering
  \includegraphics[scale=0.55]{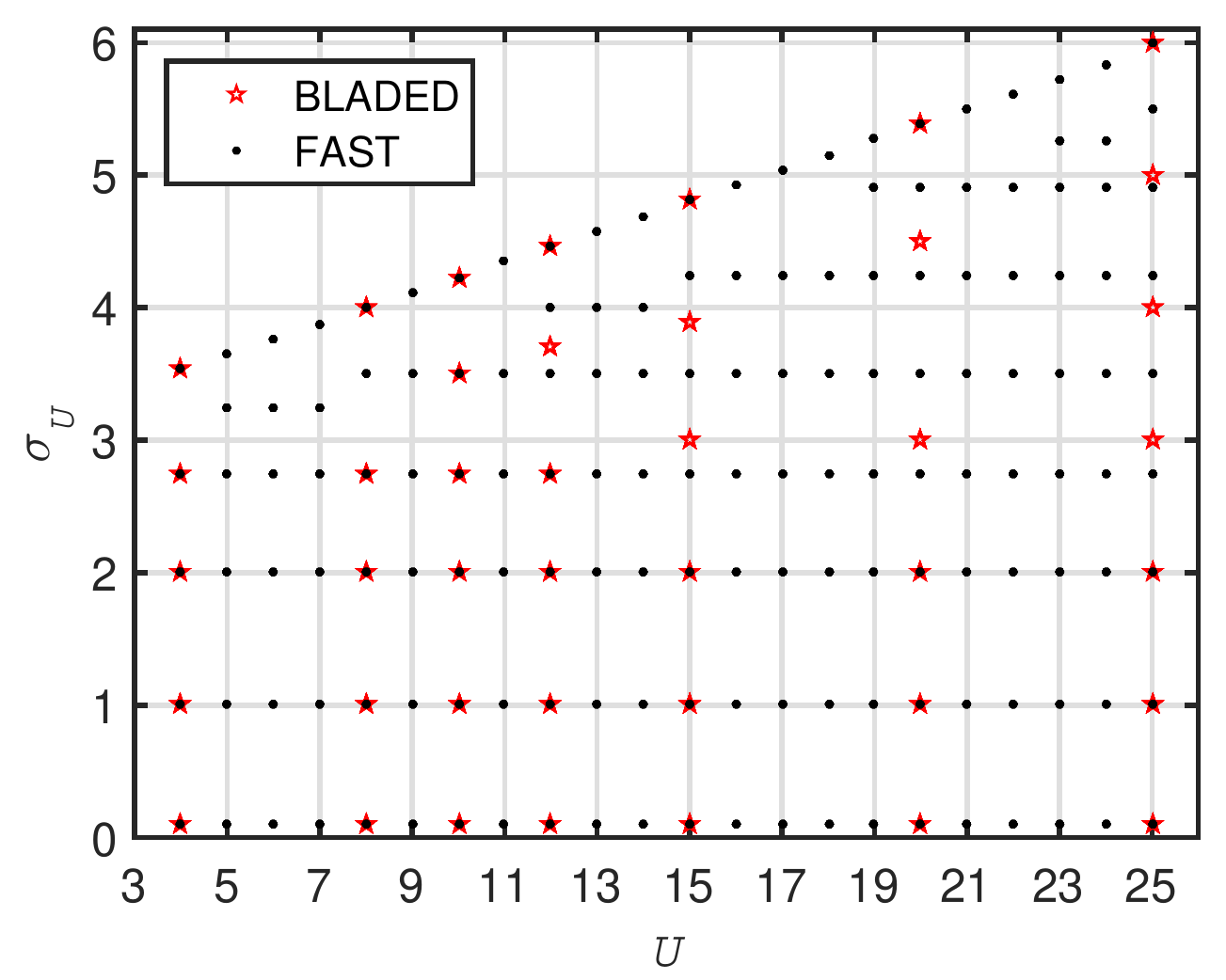}
\caption{}
  \label{fig:DOE1}
    \vspace{0ex}
  \end{subfigure}
  \begin{subfigure}[b]{0.5\linewidth}
    \centering
  \includegraphics[scale=0.55]{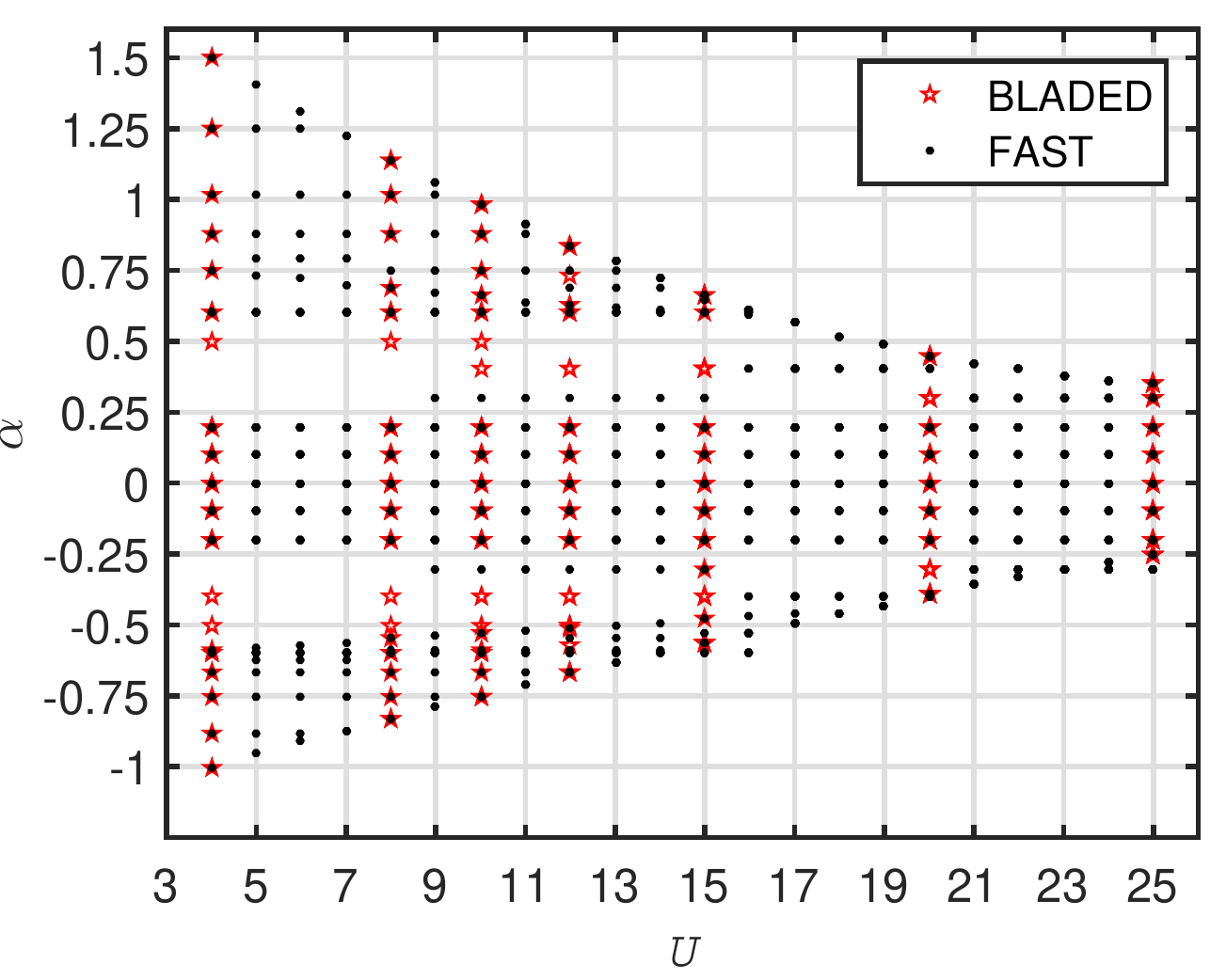}
\caption{}
  \label{fig:DOE2}
    \vspace{0ex}
  \end{subfigure} 
  \begin{subfigure}[b]{1\linewidth}
    \centering
  \includegraphics[scale=0.55]{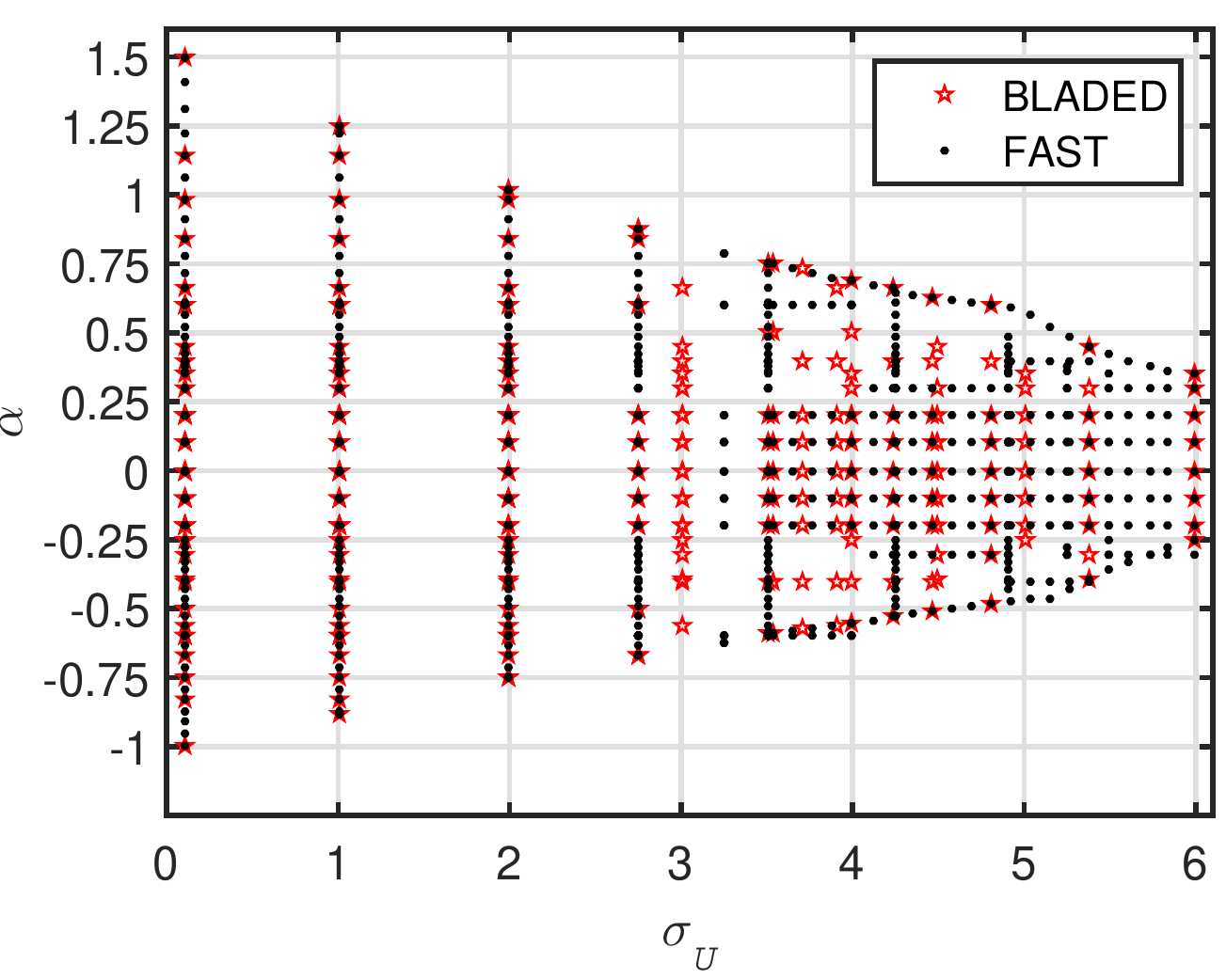}
\caption{}
  \label{fig:DOE3}
  \end{subfigure}
  \caption{Design of experiments for the FAST and Bladed simulations: (a) turbulence $\sigma_U$ as a function of mean wind velocity $U$, (b)  wind shear exponent $\alpha$ as a function of mean wind velocity $U$ and (c) shear exponent $\alpha$ as a function of Turbulence $\sigma_U$.}
  \label{fig:DOE} 
\end{figure}

The variation in the extreme loads on a wind turbine are significantly dependent on the turbulence inherent in the wind field as well as factors such as the wind shear, the mean wind velocity and the response of the turbine control system \citep{Dimitrov2015}. Turbulence intensity is given by $TI=\sfrac{\sigma_U}{U}$ where $\sigma_U$ is the standard deviation (or turbulence) and $U$ is the mean of a 10-minute wind speed time series. The wind profile above ground level is expressed using the power law relationship, which defines the mean wind velocity $U$ at a height $Z$ above ground using the wind velocity $U_{hub}$ measured at hub height $Z_h$ as reference:
\begin{equation}
	\frac{U}{U_{hub}}=\left(\frac{Z}{Z_h}\right)^\alpha
\end{equation} 
where $\alpha$ is a constant called the \emph{shear exponent}. A full factorial DOE on a non-uniform grid is produced as shown in Figure~\ref{fig:DOE} in order to examine the effects of wind velocity, inflow turbulence and shear on the simulated extreme loads from FAST and Bladed. For each combination of wind velocity, turbulence level and shear exponent we generate realizations of wind time series with 12 and 24 stochastic seeds for Bladed and FAST, respectively. However, some of the wind velocity, turbulence and shear exponent combinations are excluded because they are un-physical, thus resulting in a total of $4,344$ and $33,480$ 10-minute time series simulations for Bladed and FAST, respectively. As shown in Figure~\ref{fig:DOE} the low- and high-fidelity simulators are not sampled at exactly the same input locations. One 10-minute time series simulation using FAST takes approximately five minutes in real time, and approximately 30 minutes using Bladed, hence the much smaller number of the Bladed simulations compared to the FAST simulations. The output used are the maxima of the  blade root flapwise bending moment extracted for each of the 10-minute time series.
All low fidelity simulations are used as training samples to build the low fidelity Kriging surrogate model. A subset of the high-fidelity simulations corresponding to the blade root flapwise extreme bending moments at only three wind velocities $V=\left[4,10,25\right]m/s$ are used as a training sample to build the Hierarchical Kriging surrogate model while the remaining outputs are used as validation points.

\subsection{Stochastic Aero-servo-elastic simulations}
FAST \cite{Jonkman2005} is a wind-turbine-specific time domain aero-servo-elastic computer simulator that employs a combined modal and multibody dynamics formulation, and has limited degrees of freedom (DOF). Since FAST models flexible elements using a modal representation, the reliability of this representation depends on the generation of accurate mode shapes by the engineer, which are then used as input into FAST. Large structural elements such as blades and tower models use properties such as stiffness and mass per unit length to represent the flexibility characteristics \citep{Jonkman2005}. FAST models the turbine using 24 DOF, including two blade-flap modes and one blade-edge mode per blade, two fore-aft and two side-to-side tower bending modes, nacelle yaw, the generator azimuth angle and the compliance in the drive train between the generator and hub/rotor.  The aerodynamic model is based on the Blade Element Momentum theory (including skew inflow, dynamic stall and generalized dynamic wake).  The stochastic wind field used the Kaimal turbulence model \citep{Kaimal1972}.  Bladed \cite{Bladedtheory2003} is also a wind-turbine-specific time domain aero-servo-elastic computer simulator. The structural dynamics within Bladed are based on a modal and FEM formulation (geometric non-linear by use of a co-rotational formulation). The blade is modelled using up to 12 modes, six blade-flap and six blade-edge per blade. It also has three fore-aft and three side-to-side tower bending modes in addition to nacelle yaw. Gearbox and drivetrain dynamics are included. The aerodynamic model is also based on the Blade Element Momentum theory (including skew inflow, dynamic stall and generalized dynamic wake) \citep{HANSEN2001}. The stochastic wind field used the Mann turbulence model \citep{Mann1998}. Both the FAST and Bladed aero-servo-elastic simulations were performed with exactly the same basic control systems in the form of an external Dynamic-Link library (or DLL). The FAST and Bladed simulations did not use exactly the same input parameters in the structural and aerodynamic sub-models. For example in Bladed we use the Mann turbulence model whereas the Kaimal turbulence model is used in FAST. This illustrates the fact that different options can be used by different users in such complex simulations.

\begin{figure}[h!]
  \begin{subfigure}[b]{1\linewidth}
    \centering
  \includegraphics[scale=0.45]{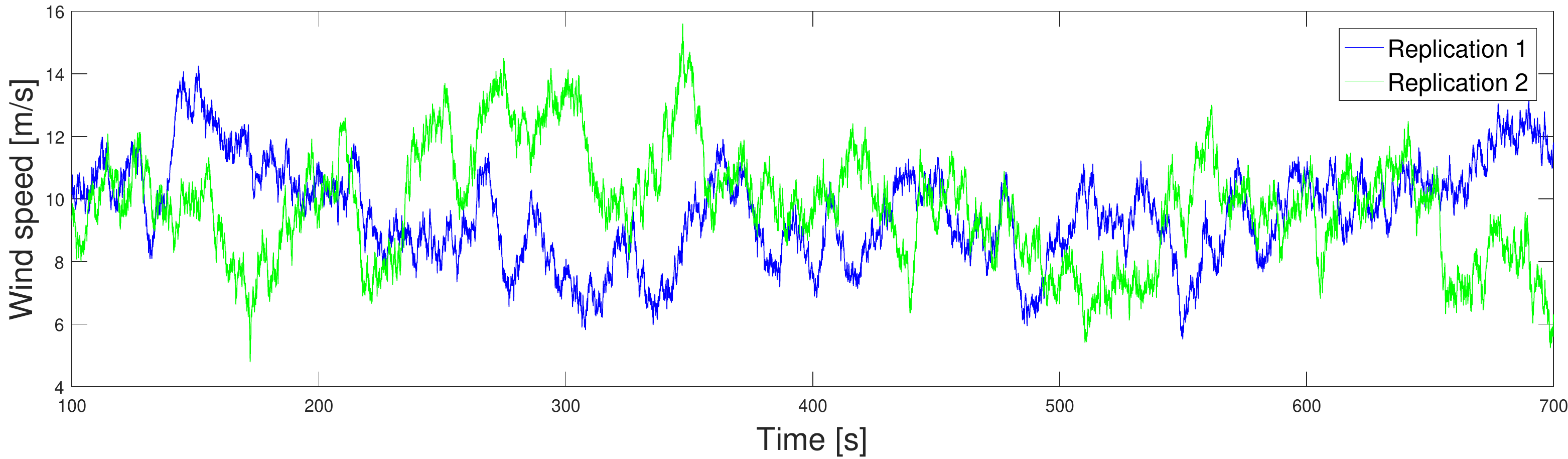}
\caption{}
    \vspace{0ex}
  \end{subfigure}
  \\
  \begin{subfigure}[b]{1\linewidth}
    \centering 
\includegraphics[scale=0.45]{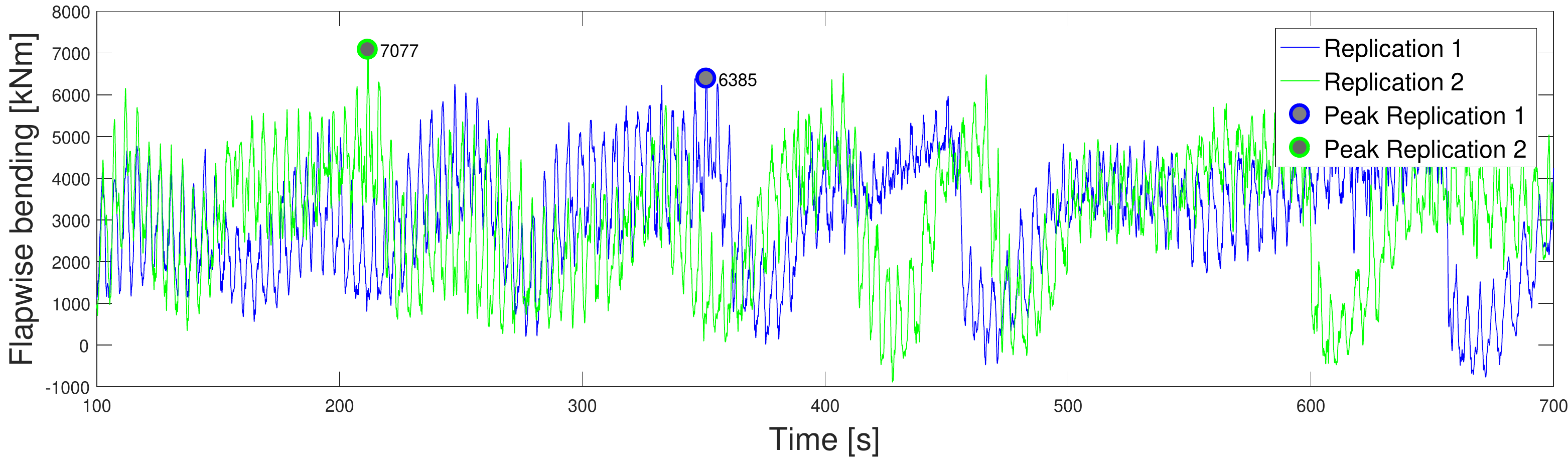}
  \caption{}
  \end{subfigure}
  \caption{(a) Time series of the wind speed. In both replications the mean wind velocity is $10~m/s$ and turbulence is $1.8~m/s$. (b) Time series of the blade root flapwise bending moment. The peak response differs in replication \#1 versus replication \#2.}
  \label{fig:lreplications}
\end{figure}

\begin{figure}[h!]
  \begin{subfigure}[b]{1\linewidth}
    \centering
  \includegraphics[scale=0.51]{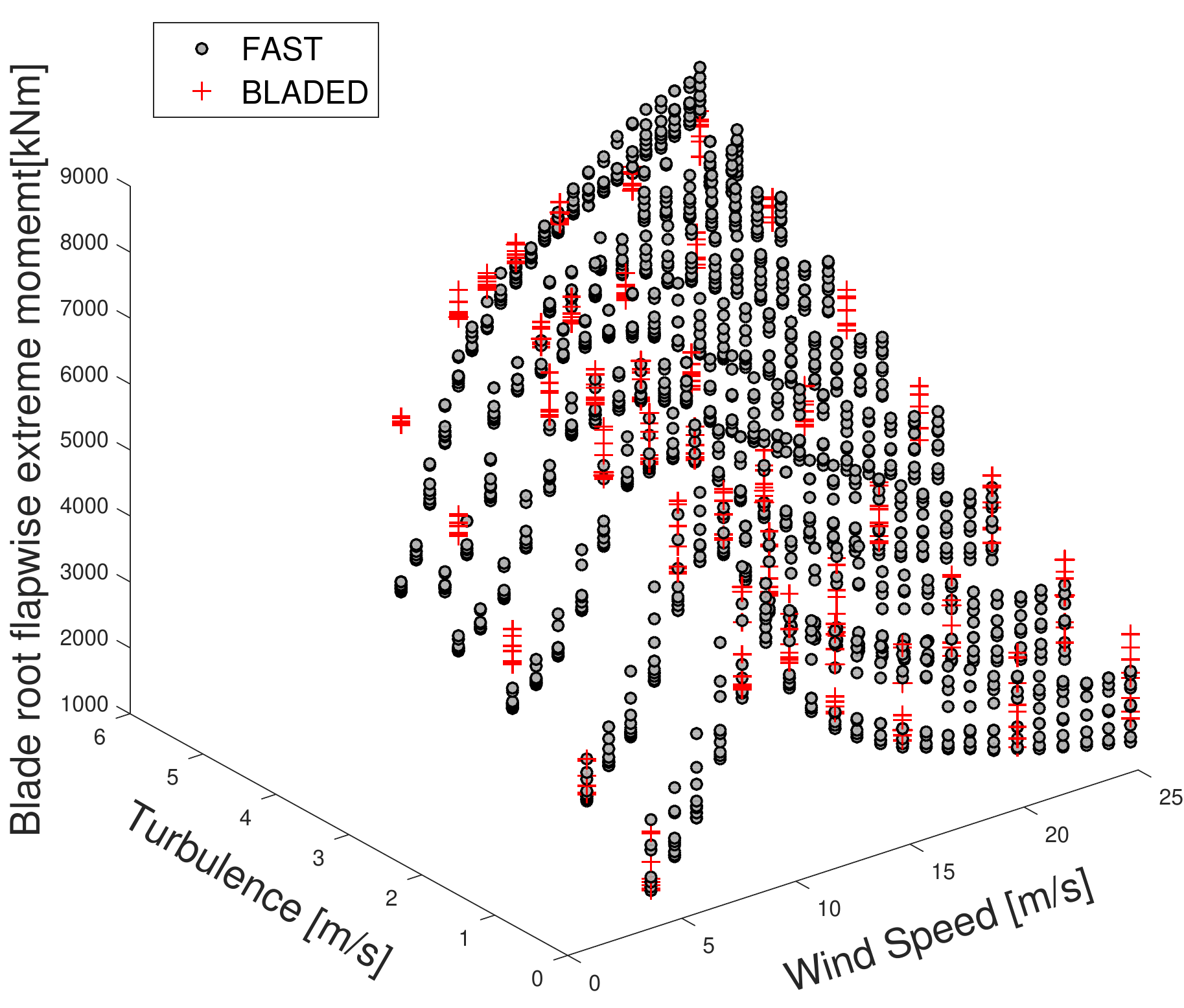}
  \label{fig:scatterblade}
  \end{subfigure}
  \caption{Scatter plot of the blade root extreme flapwise bending moment as a function of wind velocity $U$ and turbulence $\sigma_U$ (for all wind shear exponents $\alpha$). Each dot and cross are the mean of 24 and 12 replications, respectively.}
  \label{fig:scatterloads} 
\end{figure}

Aeroelastic simulations of wind turbines are stochastic due in large part to the stochastic nature of the simulated wind field and to a lesser but not insignificant degree due to the response of the control system. Figure~\ref{fig:lreplications}(a) shows two replications of the wind speed time series with mean value of $U=10~m/s$ and standard deviation (turbulence) of $\sigma_U=1.8~m/s$. The corresponding output times series of the blade root flapwise bending moment of the wind turbine is shown in Figure~\ref{fig:lreplications}(b); We observe that given the same mean wind velocity and standard deviation, the peak bending moment in replication \#1 differs from the peak bending moment in replication \#2 by approximately $10\%$. When repeating such aero-servo-elastic simulations for a range of wind speeds ($U$), turbulence ($\sigma_U$) and shear exponents ($\alpha$) in the DOE, and extracting the peak from each time series we get a representation of the blade root extreme flapwise bending moment as shown in Figure~\ref{fig:scatterloads}. This is the data that will be approximated with Kriging surrogate models. Figure~\ref{fig:scatterloads} demonstrates how the magnitude and scatter (noise) of the replications changes as a function of $U$, $\sigma_U$ and $\alpha$. Note how the magnitude of the blade root extreme flapwise bending moment differs between FAST and Bladed. However, they display similar trends.

\subsection{Comparing Hierarchical Kriging and conventional Kriging}  \label{sec: compoptkrigmodels}
\begin{figure}[!ht]
  \begin{subfigure}[b]{0.5\linewidth}
	   \centering
  		\includegraphics[scale=0.55]{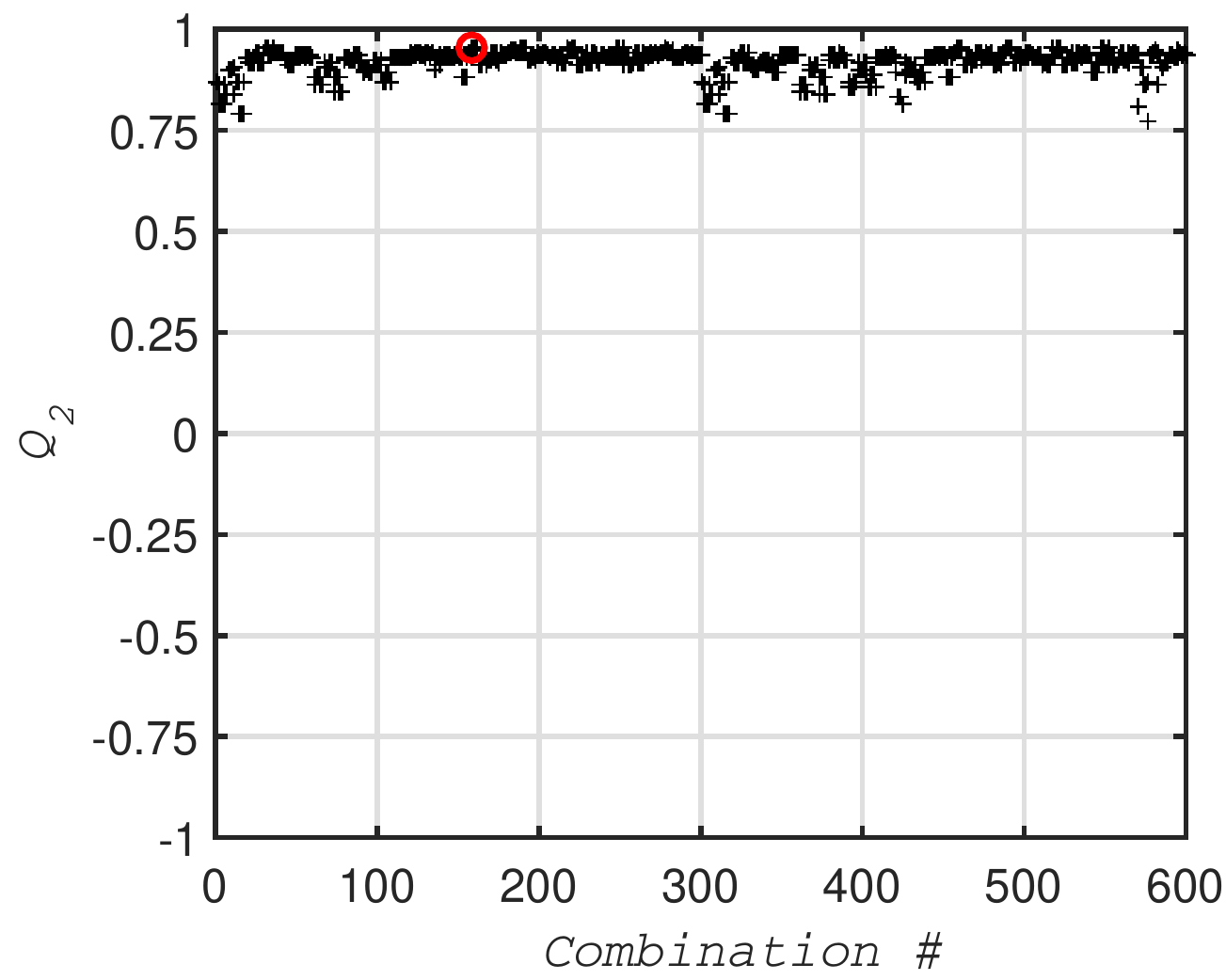}
  \end{subfigure}
  \begin{subfigure}[b]{0.5\linewidth}
	   \centering
  		\includegraphics[scale=0.55]{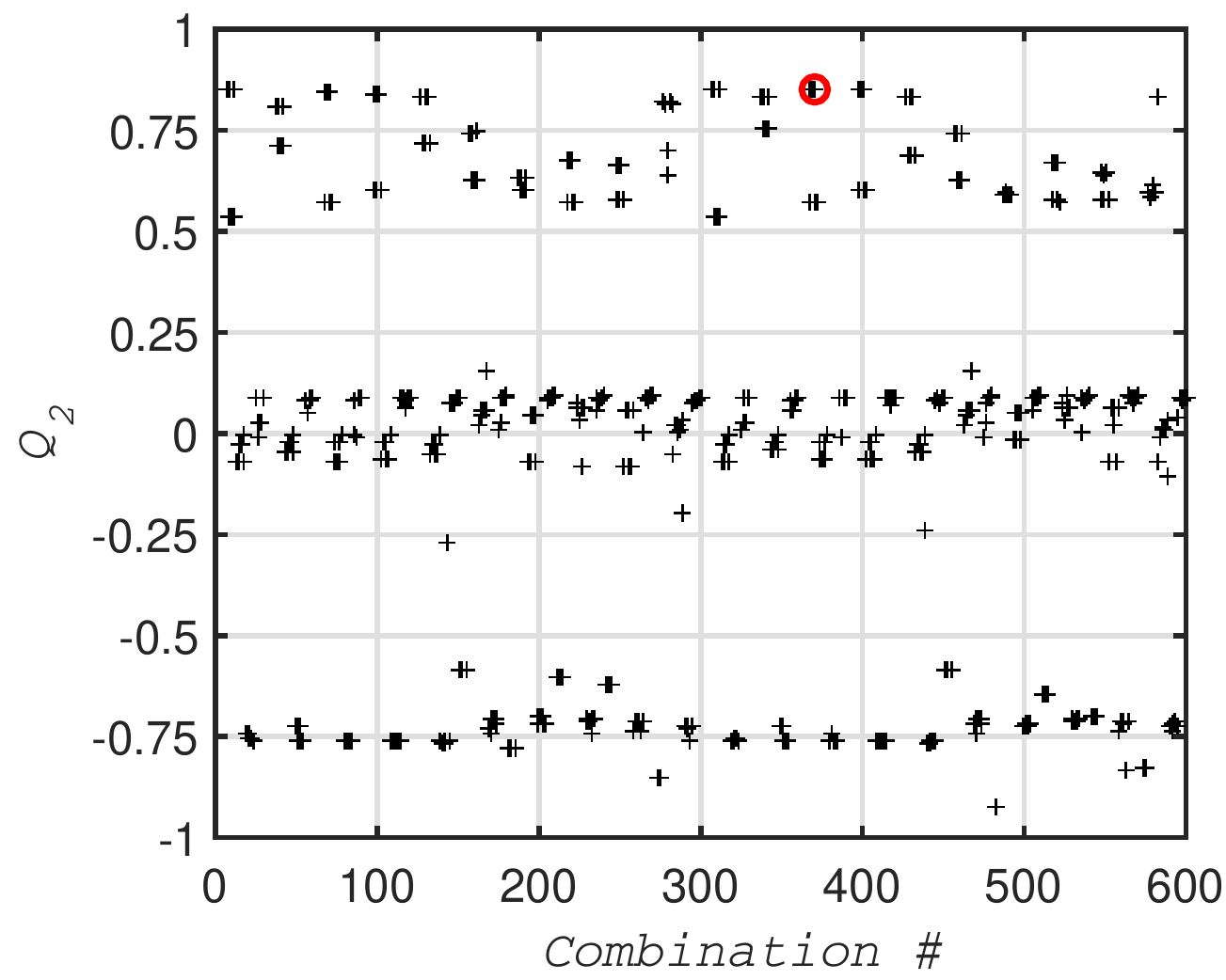}
  \end{subfigure}\\
  \begin{subfigure}[b]{0.5\linewidth}
	   \centering
  		\includegraphics[scale=0.55]{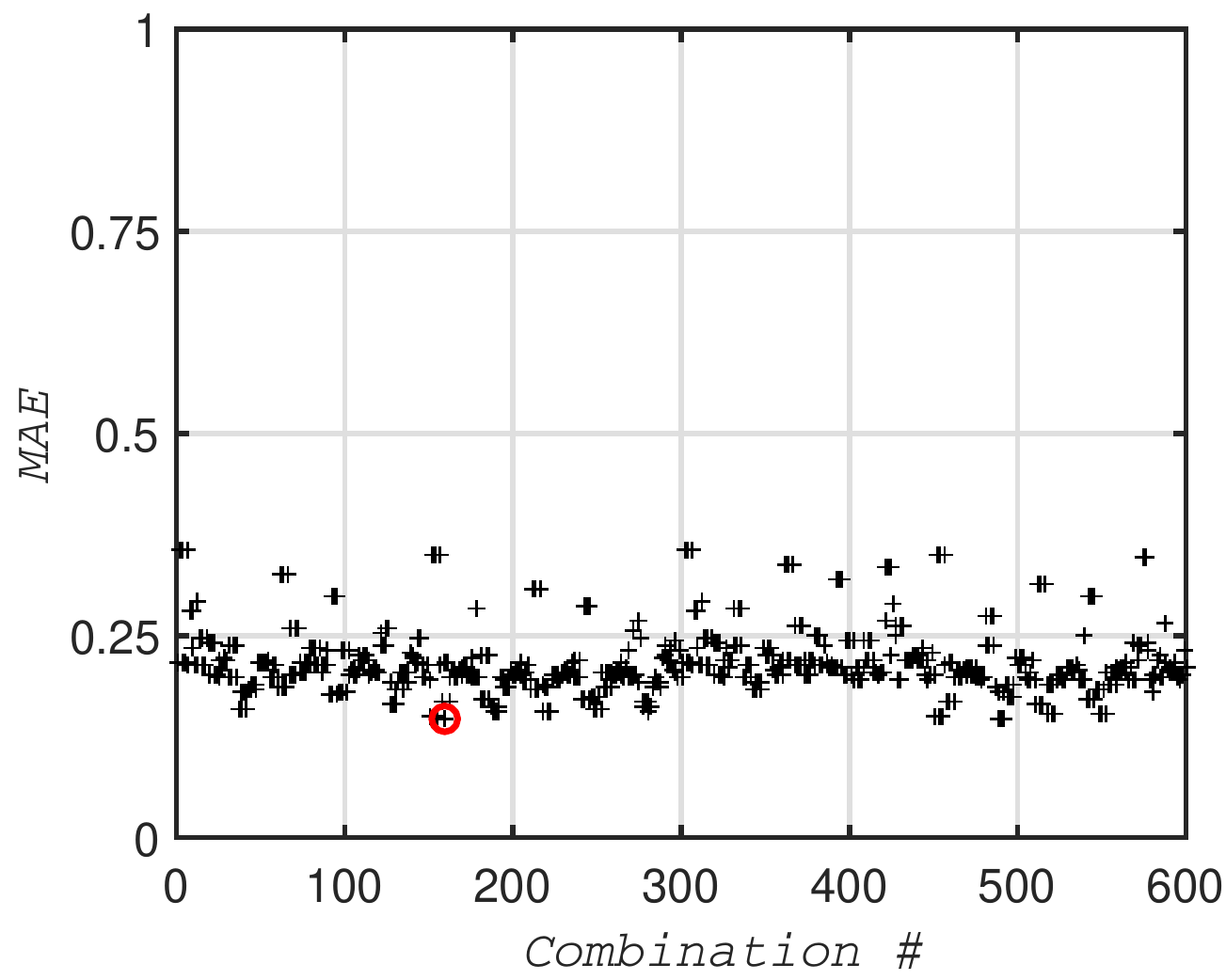}
  		\caption{Hierarchical Kriging}
  \end{subfigure}
  \begin{subfigure}[b]{0.5\linewidth}
	   \centering
  		\includegraphics[scale=0.55]{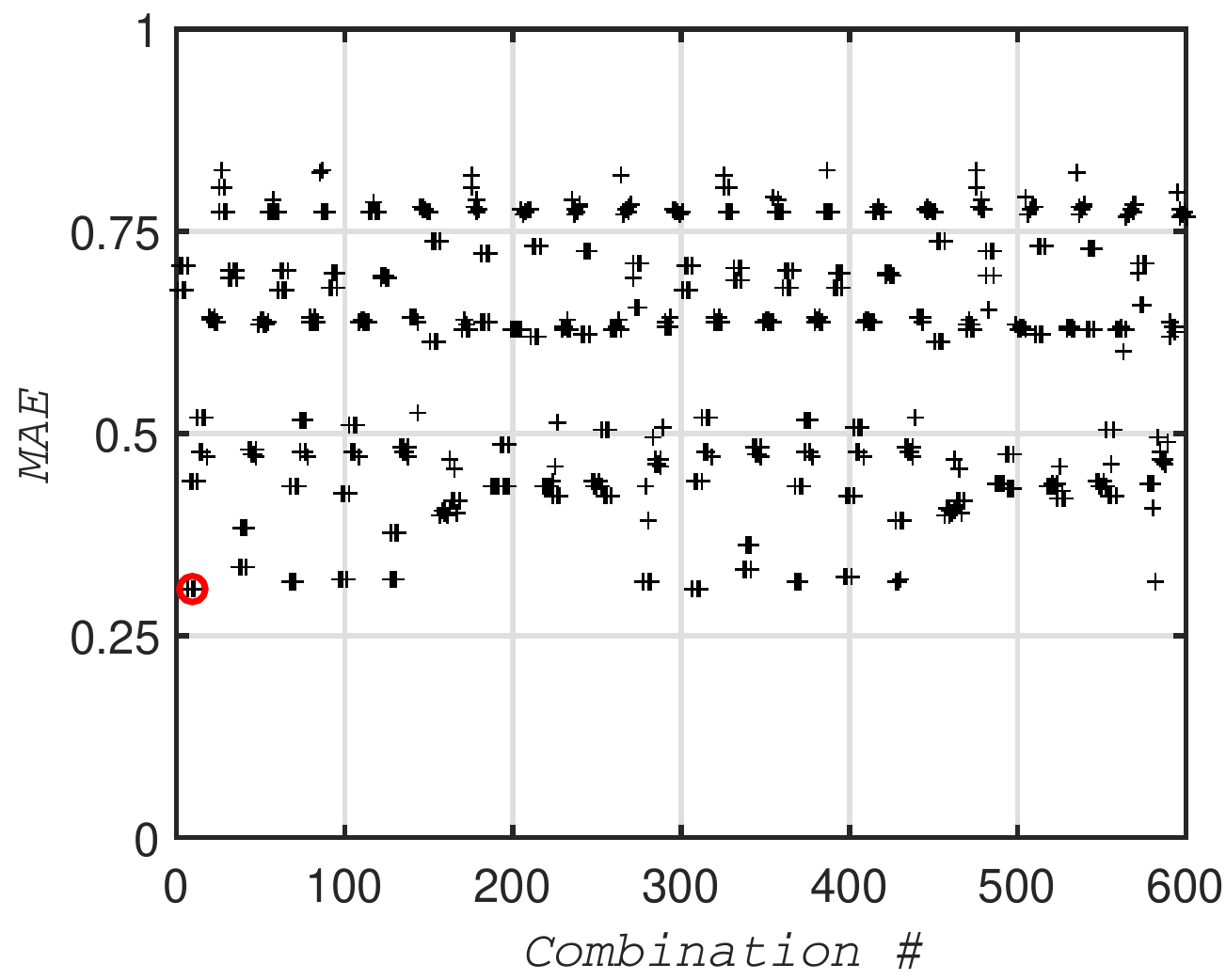}
  		\caption{Conventional Kriging}
  \end{subfigure}
  		\caption{Surrogate models estimation errors $Q_2$ and $MAE$ as a function of the combination numbers of the Kriging parameters. The left column contains the estimation error measures for Hierarchical Kriging and the right column contains the estimation error measures for conventional Kriging of the high-fidelity simulator outputs. The red circles represents the combination of the Kriging parameters resulting in the best surrogate models.}
  		  \label{fig:valerrormeasures}
\end{figure}

In Figure~\ref{fig:valerrormeasures} the estimation errors $Q_2$ and $MAE$ of the high fidelity simulations are plotted as a function of the 600 combinations of the Kriging parameters for both the Hierarchical Kriging and conventional Kriging. We first observe that Hierarchical Kriging generally yields lower normalized $MAE$ and higher $Q_2$ compared to conventional Kriging of the high-fidelity simulations. We also observe that the estimation errors of Hierarchical Kriging vary within a limited range for $Q_2 \in [0.7699,0.9571]$ and $MAE \in [0.1474,0.3569]$ compared to conventional Kriging whose estimation errors vary over significantly larger range for $Q_2 \in [-0.9221,0.8544]$ and $MAE \in [0.3077,0.8251]$. This indicates that Hierarchical Kriging is far less sensitive to the choice of the Kriging parameters compared to conventional Kriging. We note that for conventional Kriging, certain models show negative $Q_2$ values, which indicates extremely poor surrogate models. In Figure~\ref{fig:valerrormeasures}, all combinations of the Kriging parameters resulted in a Hierarchical Kriging surrogate models whose $Q_2$ error measure is larger than 0.76, when in fact only 38 out of 600 combinations are within this range for the conventional Kriging surrogate models of the high-fidelity simulations. The largest $MAE$ of the Hierarchical Kriging surrogate models is 0.3569 and is roughly equal to the smallest $MAE$ of the conventional Kriging. This shows that Hierarchical Kriging is more accurate compared to conventional Kriging in this real case application. Furthermore, from Table~\ref{tab:paramstudyresults} we observe that the two estimation error measures consistently predict that combination \#160 of the Kriging parameters yields the best Hierarchical Kriging surrogate model. In contrast, for conventional Kriging $Q_2$ and $MAE$ predict that combination \#371 and \#9 yield the best surrogate model, respectively. This indicates that Hierarchical Kriging is potentially far less sensitive to the choice of the estimation error measure when evaluating the surrogate models compared to conventional Kriging. The details of the Kriging parameters that yield the best Hierarchical Kriging and conventional Kriging surrogate models are shown in Table~\ref{tab:bestoptionsHirarKriDirecKri}. The above observations indicate that in the presence of a limited number of high-fidelity noisy simulations output, Hierarchical Kriging tends to be more accurate, more stable and more predictable compared to conventional Kriging.

\begin{figure}[t]
  \begin{subfigure}[b]{0.5\linewidth}
	   \centering
  		\includegraphics[scale=0.55]{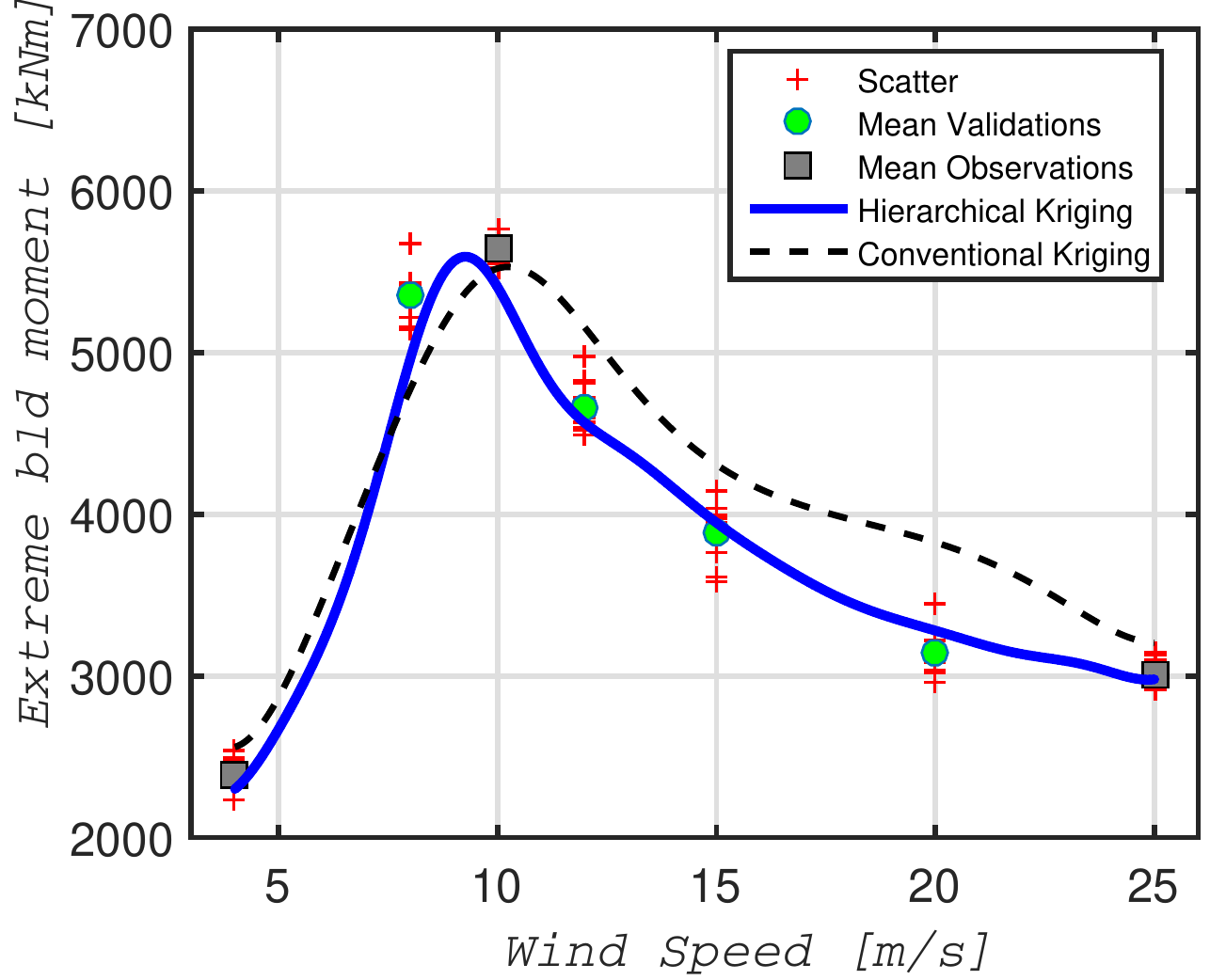}
  		\caption{}
   \end{subfigure}
  \begin{subfigure}[b]{0.5\linewidth}
	   \centering
  		\includegraphics[scale=0.55]{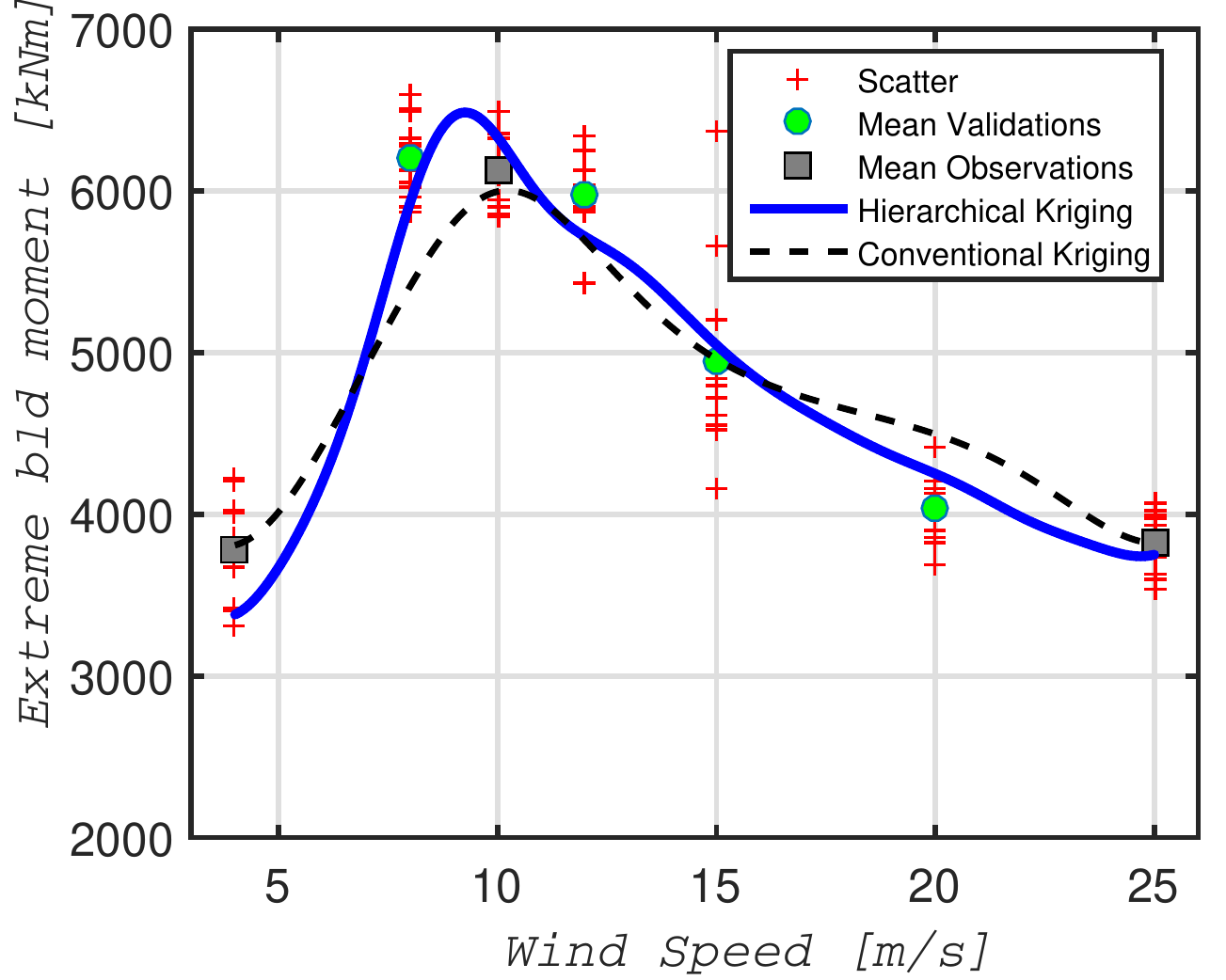} 
  		\caption{}
  \end{subfigure} \\
  \begin{subfigure}[b]{0.5\linewidth}
	   \centering
  		\includegraphics[scale=0.55]{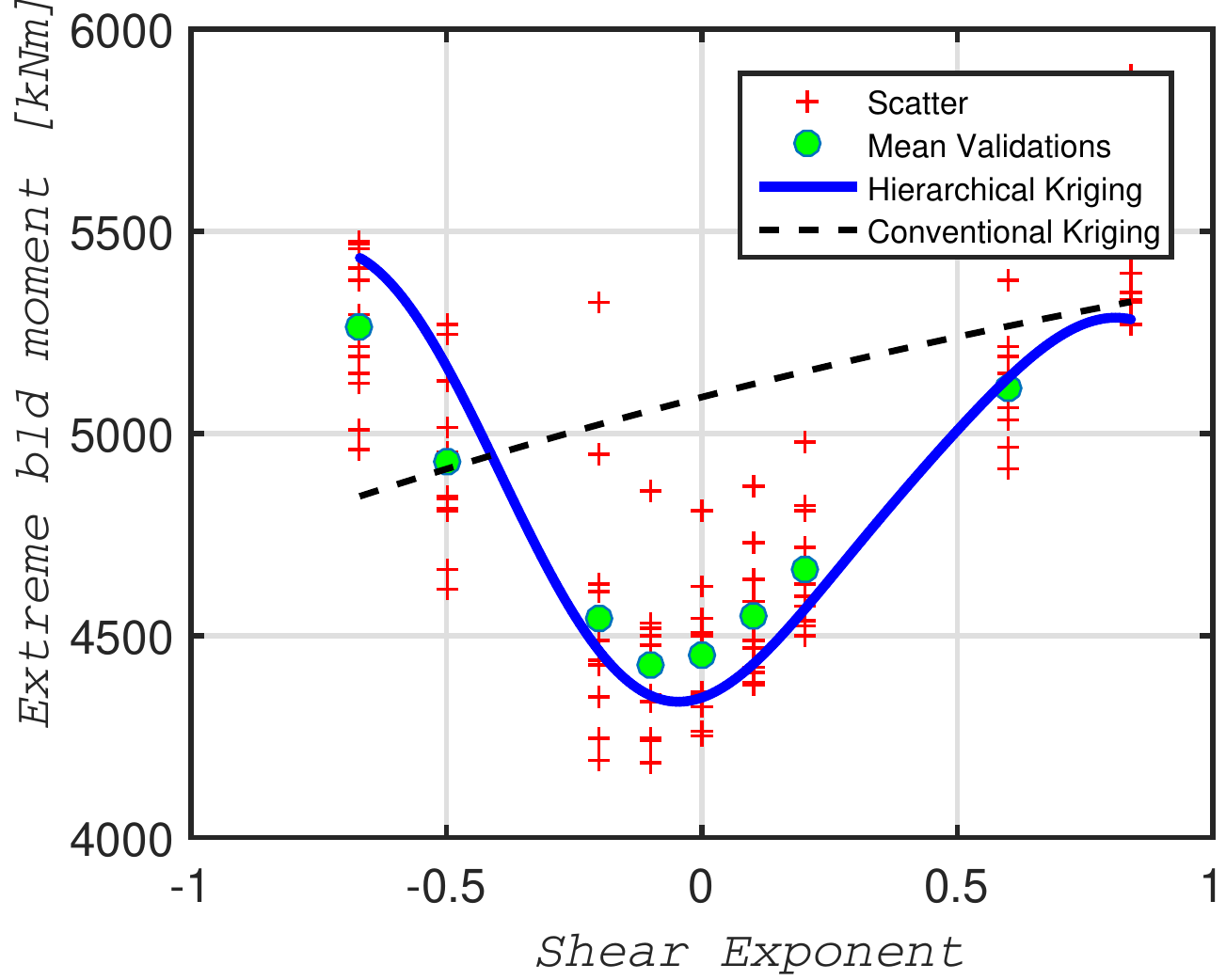}
  		\caption{}
  \end{subfigure}
  \begin{subfigure}[b]{0.5\linewidth}
	   \centering
  		\includegraphics[scale=0.55]{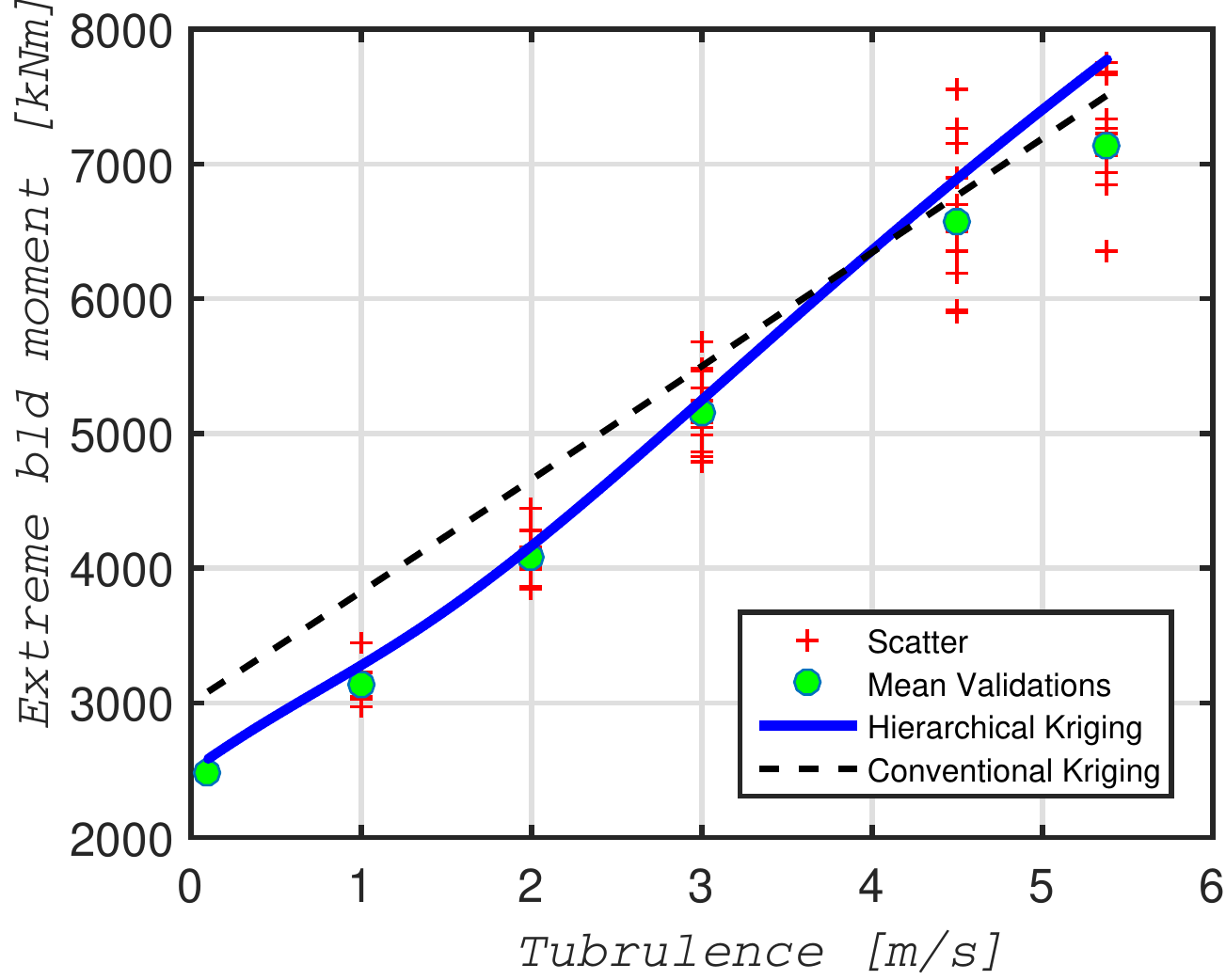}
  		\caption{}
  \end{subfigure}
  		\caption{Comparison of the best Hierarchical Kriging  and best conventional Kriging surrogate models of the high-fidelity extreme blade root flapwise bending moment. The best Hierarchical Kriging and best conventional Kriging surrogate models are selected based on the $Q_2$ estimation error measure (see Table~\ref{tab:bestoptionsHirarKriDirecKri}). The input space is sliced as follows: (a) $\sigma_U=1.06~m/s$ and $\alpha=0.2$, (b) $\sigma_U=2.0~m/s$ and $\alpha=-0.2$, (c) $U=12~m/s$ and $\sigma_U=~1.0m/s$, (d) $U=~20m/s$ and $\alpha=~0.2$.}
  		    \label{fig:slicecompareBest}
\end{figure}

For an easy visualization and validation of the surrogate models, we take a slice in the three-dimensional input space and plot in Figure~\ref{fig:slicecompareBest} the best Hierarchical Kriging surrogate model and the best conventional Kriging model of the high-fidelity blade root extreme flapwise bending moment. Qualitatively we see that the best  Hierarchical Kriging and the best conventional Kriging surrogate models predictions are both close to the high-fidelity validation outputs except for the extreme loads as a function of the wind shear exponent (Figure~\ref{fig:slicecompareBest} (c)) where only Hierarchical Kriging is able to capture the trend of the response. The results in Figure~\ref{fig:slicecompareBest} indicate that given a DOE, a broad search through all the Kriging parameters may indeed happen to yield an accurate conventional Kriging surrogate model of the high-fidelity simulations at par with Hierarchical Kriging. However, a practical user of conventional Kriging may not perform a broad search and will opt instead for a set of Kriging parameters based on former experiences, assumptions and engineering judgements that will indeed yield an inaccurate conventional Kriging surrogate model most of the time as shown in the estimation error plots (Figure~\ref{fig:valerrormeasures}). Conversely, Hierarchical Kriging is more robust and will yield an accurate surrogate model consistently most of the time regardless of the Kriging parameters used.  According to Table~\ref{tab:bestoptionsHirarKriDirecKri} the best conventional Kriging surrogate model of the high-fidelity simulations has an ellipsoidal and isotropic Mat\'ern-5/2 correlation kernel. However, a practical user may instead opt for an anisotropic exponential correlation kernel with a second-order polynomial trend, as shown in Table~\ref{tab:options495DirecKri}. 

\begin{figure}[!ht]
  \begin{subfigure}[b]{0.5\linewidth}
	   \centering
  		\includegraphics[scale=0.55]{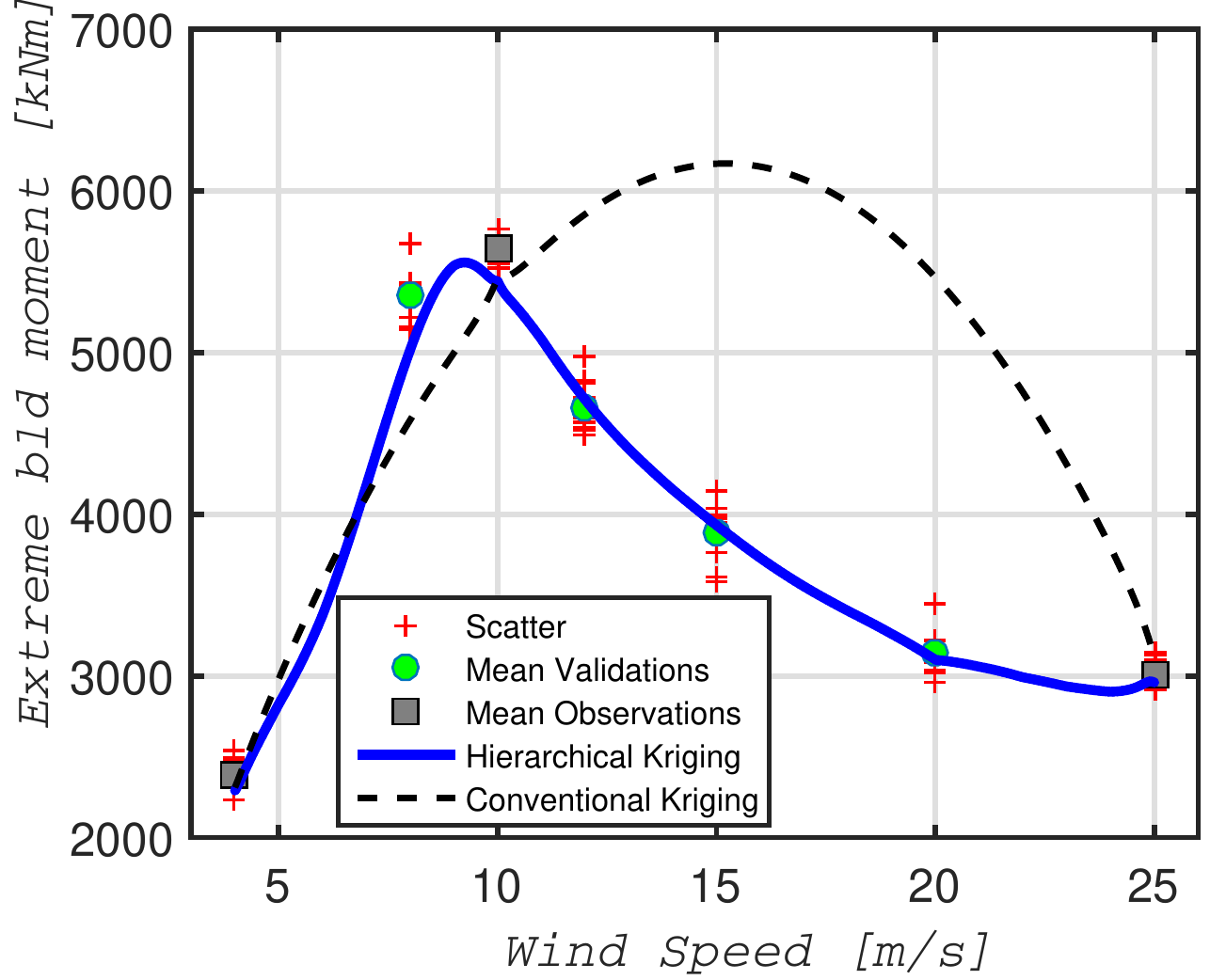}
  		\caption{}
   \end{subfigure}
  \begin{subfigure}[b]{0.5\linewidth}
	   \centering
  		\includegraphics[scale=0.55]{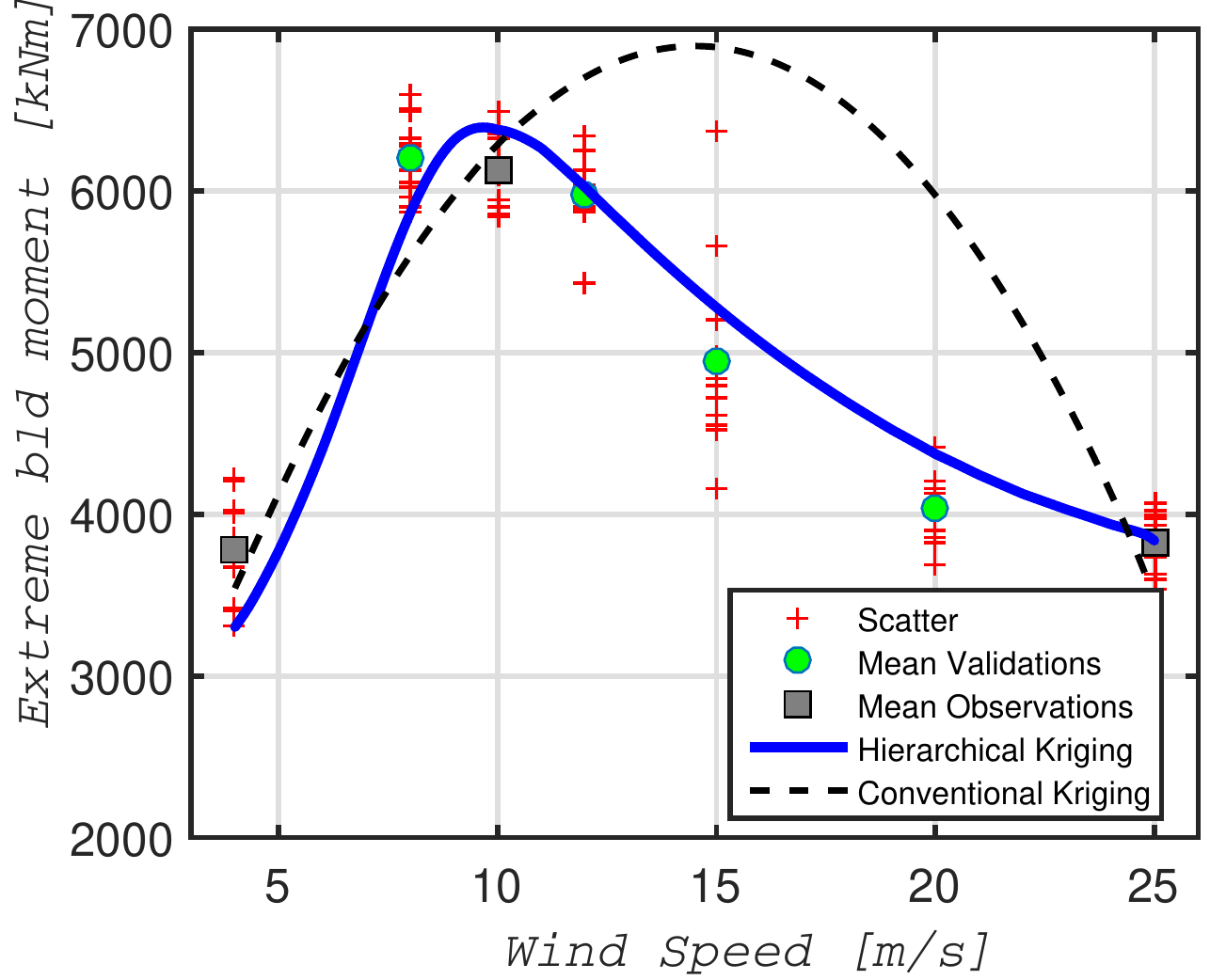}
  		\caption{}
  \end{subfigure} \\
  \begin{subfigure}[b]{0.5\linewidth}
	   \centering
  		\includegraphics[scale=0.55]{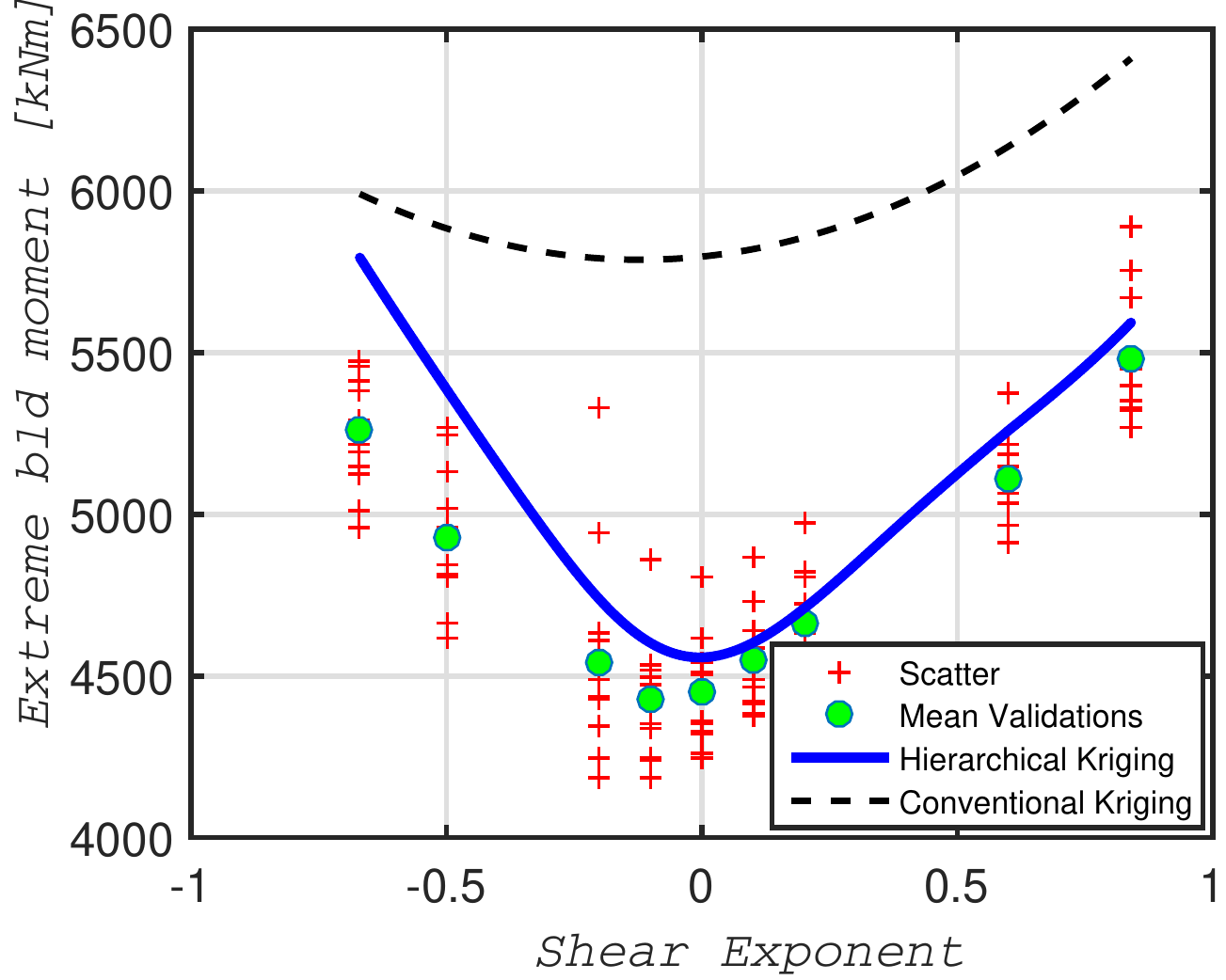}
  		\caption{}
  \end{subfigure}
  \begin{subfigure}[b]{0.5\linewidth}
	   \centering
  		\includegraphics[scale=0.55]{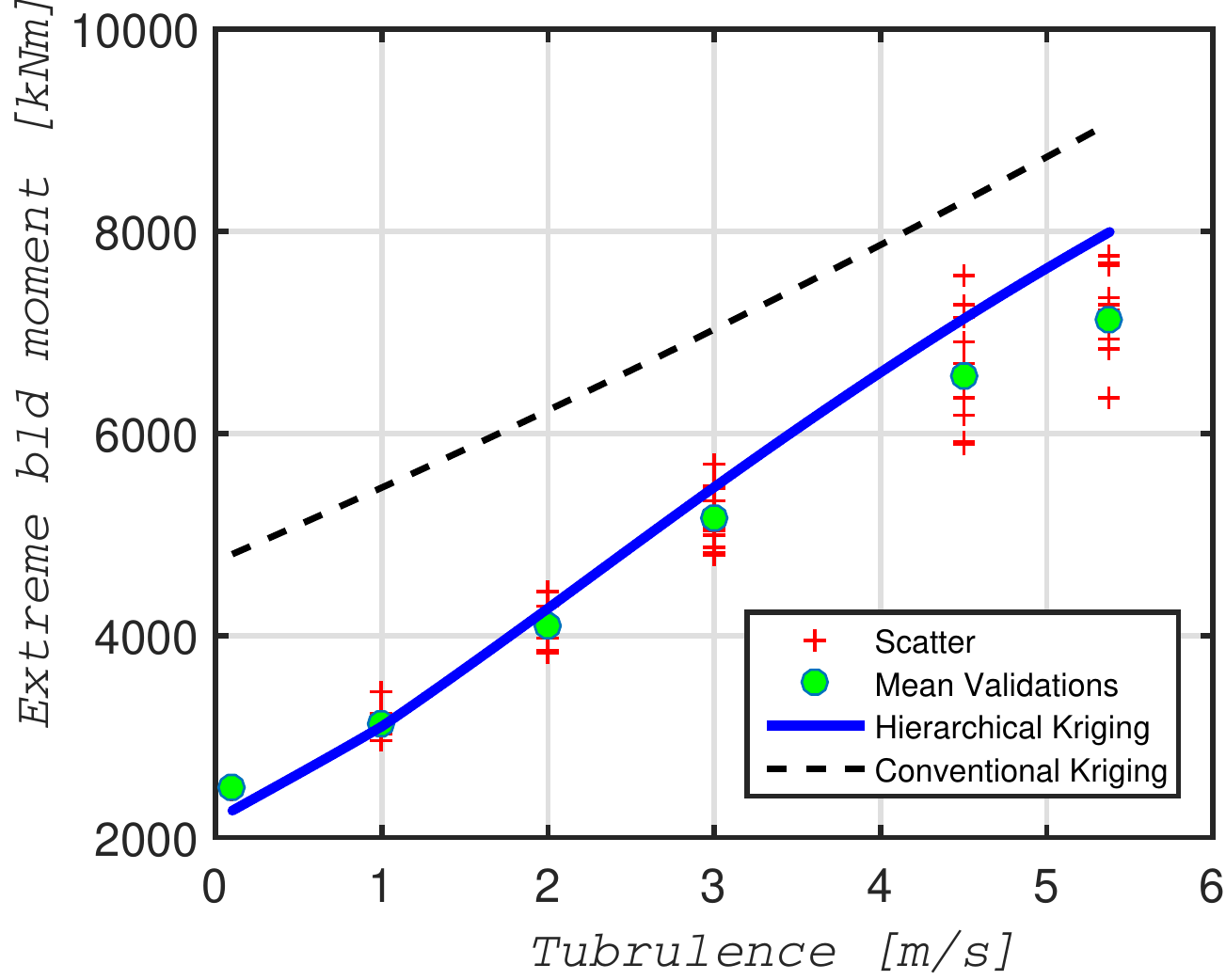}
  		\caption{}
  \end{subfigure}
  		\caption{Comparison of the Hierarchical Kriging and conventional Kriging surrogate models of the high-fidelity extreme blade root flapwise bending moment, based on the Kriging parameters in combination \#495 (see Table~\ref{tab:options495DirecKri}). The input space is sliced as follows: (a) $\sigma_U=1.0~m/s$ and $\alpha=0.2$, (b) $\sigma_U=2.0~m/s$ and $\alpha=-0.2$, (c) $U=12~m/s$ and $\sigma_U=1.0~m/s$, (d) $U=20~m/s$ and $\alpha=0.2$}
  		    \label{fig:slicecompareBesttorandom}
\end{figure}

In Figure~\ref{fig:slicecompareBesttorandom} we plot the Hierarchical Kriging and conventional Kriging surrogate models corresponding to the Kriging parameters in Table~\ref{tab:options495DirecKri} that are selected from engineering judgement, i.e. not optimized. This plot showcases how using the low-fidelity Kriging model as a trend significantly improves the predictive accuracy of the Hierarchical Kriging model compared to conventional Kriging for a given set of sound Kriging parameters. Conventional Kriging gives a poor approximation of the high-fidelity validation points, while Hierarchical Kriging performs notably better. Finally we conclude that, for most practical problems where the best Kriging parameters are not known beforehand and where the trend of the output as a function of the various input dimensions are not known either, additional test data are not necessarily available to validate the surrogate models and a broad search for the best Kriging parameters is not necessarily feasible. In this case the use of Hierarchical Kriging surrogate modelling for limited number of high-fidelity simulations may prove a robust and consistent method regardless of the choice of the Kriging parameters.

\begin{table}[!ht]
\scriptsize
  \caption{Combination numbers of the Kriging parameters that resulted in the best Hierarchical Kriging and conventional Kriging surrogate models of the high-fidelity simulations and their corresponding estimation error values.}
  \label{tab:paramstudyresults}
  \centering
\begin{tabular}{Q L L L L}
      \hhline{=====}
      \multirow{2}{*}{Error metric}         & \multicolumn{2}{c}{\underline{Hierarchical Kriging}} &  \multicolumn{2}{c}{\underline{Conventional Kriging}} \\
      								                                 &  Best Combination \# & Error metric value   &Best Combination \# & Error metric value      \\     
      \midrule
      $Q_2$					& 160 &  0.9571        & 371   &0.8544 \\
      $MAE$					& 160 &  0.1474        & 9       &0.3077 \\
      \hhline{=====}
\end{tabular}
\end{table}

\begin{table}[!ht]
\scriptsize
  \caption{Kriging parameters that yield the best Hierarchical Kriging and conventional Kriging surrogate models of the high-fidelity blade root extreme flapwise bending moment.}
  \label{tab:bestoptionsHirarKriDirecKri}
  \centering
\begin{tabular}{p{2.0cm} T T T  T  p{2.0cm}  p{2.0cm} }
      \hhline{=======}
      & Correlation type & Correlation family &  Correlation isotropy & Trend type & hyper-parameters estimation method  & hyper-parameters optimization method\\
          \midrule
Best Hierarchical Kriging (combination \#160)  &  Separable 		& Gaussian       & False 		    &		1st order poly.							 & MLE		    & HGA				\\
          \hdashline
	\\
Best conventional Kriging  (combination \#371) &  Ellipsoidal 		& Mat\'ern-5/2       & True 		    &		1st order poly.							 & CV		    & HSADE				\\
      \hhline{=======}
\end{tabular}
\end{table}

\begin{table}[!ht]
\scriptsize
  \caption{Kriging parameters chosen based on engineering judgement. The corresponding Hierarchical Kriging and conventional Kriging surrogate models are plotted in Figure~\ref{fig:slicecompareBesttorandom}.}
  \label{tab:options495DirecKri}
  \centering
\begin{tabular}{T T T T  T  p{2.9cm}  p{2.9cm} }
      \hhline{=======}
      & Correlation type & Correlation family &  Correlation isotropy & Trend type & hyper-parameters estimation method  & hyper-parameters optimization method\\
          \midrule
Kriging combination \# 495  & Ellipsoidal 		& Exponential       & False 		    &		2nd order poly.			& CV		    & HGA				\\
      \hhline{=======}
\end{tabular}
\end{table}

\section{Conclusions and outlook}
We presented a parametric Hierarchical Kriging methodology to fuse the noisy extreme flapwise bending moment at the blade root of a large wind turbine from a low-fidelity (FAST) and a high-fidelity (Bladed) aero-servo-elastic simulators. The low and high-fidelity aero-servo-elastic simulators of the wind turbine were implemented by two independent engineers, implying that uncertainty in the modelling and input assumptions are implicitly included thus representing a realistic case study. A DOE was formulated for both low and high-fidelity simulators as a function of wind velocity, wind turbulence and wind shear. With limited high-fidelity simulations samples, the Hierarchical Kriging surrogate model predictions compared well with the high-fidelity validation set. The notably accurate prediction performance is due to using the Kriging model of the low-fidelity simulations as a model trend for the high-fidelity Kriging model. The main assumption is that the high and low-fidelity aero-servo-elastic simulations follow similar trends, which makes the fusion of multi-fidelity simulations feasible. In this work, we stipulated that for most engineering problems, the set of Kringing parameters that yields the best surrogate model is not known a priori, and a search through an ensemble of Kriging parameters may prove beneficial to protect against a poor or lucky choice of a surrogate. Hence, the best Hierarchical Kriging and conventional Kriging surrogate models of the high-fidelity simulations were selected based on a parametric evaluation of all possible combinations of the following Kriging parameters candidates: (1) correlation type, (2) correlation family, (3) correlation isotropy, (4) trend types, (5) hyper-parameters estimation methods and (6) hyper-parameters optimization methods. The predictive coefficient and the maximum absolute errors were used as error measures to select the best surrogate models. The results indicate that a broad search through all possible combinations of the Kriging parameters may coincidently yield a rather accurate conventional Kriging surrogate model of the limited high-fidelity simulations at par with the accuracy of the Hierarchical Kriging surrogate model. However, for most practical engineering problems where the best Kriging parameters are not known a priori, test data are not necessarily available to validate the surrogate models and a broad and detailed search for the best Kriging parameters is in practice never carried out. In this context we have shown that the use of parametric Hierarchical Kriging surrogate modelling proved to be a robust and consistent method because it is generally not sensitive to the choice of the Kriging parameters nor to the choice of the estimation error measure compared to conventional Kriging. 

There are three interesting extensions to this research. The first one is to investigate if the potential gains from using Hierarchical Kriging hold in high dimensional engineering problem (e.g. 10 input variables or more). The second is to investigate in more detail which estimation error measures (or a combination thereof) are best suited for Hierarchical Kriging. Finally, in the process of parametric Hierarchical Kriging, a large number of surrogate models were generated; it would be interesting to investigate the potential improvements that an ensemble approach could bring to the predictions of Hierarchical Kriging.

\section*{Acknowledgements}
The main author gratefully acknowledges MiTa-Teknik A/S for providing the control system DLL for the FAST and BLADED aero-servo-elastic
simulations.


\bibliographystyle{chicago}
\bibliography{biblioImad}

\end{document}